\documentclass[prb,aps,
 reprint,
 amsmath,amssymb
]{revtex4-2}

\usepackage{lipsum}  
\usepackage[dvipsnames]{xcolor}
\usepackage{tikz}
\usepackage{braket}
\usepackage{graphicx}% Include figure files
\usepackage{dcolumn}% Align table columns on decimal point
\usepackage{bm}% mathbf math
\usepackage{hyperref}
\usepackage{xcolor}
\usepackage{tabularx}
\usepackage{array}
\usepackage{esint}
\usepackage{multirow}
\usepackage{minitoc}
\newcolumntype{Y}{>{\centering\arraybackslash}X}
\hypersetup{
    colorlinks=true,
    linkcolor=blue,
    filecolor=blue,      
    urlcolor=blue,
    citecolor=blue,
    }

\makeatletter
\def\bibliography#1{\relax}
\makeatother

\begin{document}

\preprint{APS/123-QED}

\title{One-dimensional topology and topolectrics of nonsymmorphic Kramers degenerate systems}

\author{Max Tymczyszyn}
\author{Edward McCann}%
 \email{ed.mccann@lancaster.ac.uk}
\affiliation{%
 Department of Physics, Lancaster University, Lancaster, LA1 4YB, United Kingdom
}%

\begin{abstract}

We describe nonsymmorphic four-band tight-binding models in one dimension with Kramers degeneracy, and propose topolectric-circuit realizations of their topological phases. We begin with a representative model in the nonsymmorphic AII class with symmorphic time-reversal symmetry and nonsymmorphic charge-conjugation and chiral symmetries, resulting in a $\mathbb{Z}_2$ invariant. We also provide a Bogoliubov-de Gennes model in the nonsymmorphic D class with symmorphic charge-conjugation symmetry and nonsymmorphic time-reversal and chiral symmetries, which supports a $\mathbb{Z}_4$ classification. To compute these invariants, we extend an open-path winding-number formulation previously used for non-Kramers-degenerate nonsymmorphic $\mathbb{Z}_2$ phases to Kramers-degenerate four-band systems. We propose topolectric circuit implementations of nonsymmorphic systems, beginning with a comparison between the symmorphic Su-Schrieffer-Heeger and nonsymmorphic charge-density-wave models. We then extend this methodology to topolectric realizations of the AII and $\mathbb{Z}_4$ models, finding that the impedance response reproduces the predicted phase boundaries and associated zero energy modes. Finally, we analyze disorder in the $\mathbb{Z}_4$ model and find that, although disorder breaks the nonsymmorphic symmetries, certain domain-wall zero modes in the minimal nearest-neighbor model remain pinned at zero energy because the disorder has no first-order coupling to the soliton subspace; longer-range terms satisfying relevant symmetries lift this emergent property.

\end{abstract}

\maketitle

%\tableofcontents

\section{Introduction}

Topological materials are a central focus of condensed matter research~\cite{kitaev01,kitaev03,sarma06,nayak08,stern13,lahtinen17,beenakker15,guo16,sato17,sharma22}, with models such as the Kitaev chain~\cite{kitaev01} guiding the pursuit of experimentally realizable nontrivial superconducting phases~\cite{dvir23}. This prospect has driven extensive theoretical~\cite{kitaev03,nayak08,stern13,lahtinen17,laubscher24} and experimental efforts on systems such as nanowires~\cite{mourik12,das12,deng16} and magnetic atom chains~\cite{nadj14,ruby15,awlak16} that are primarily investigated through scanning tunneling microscopy~\cite{nadj14,ruby15,awlak16,jack21}. However, implementing and tuning such platforms can be challenging \cite{chen11,fidkowski11,senthil15,lutchyn18}, motivating complementary approaches that capture topological band structure and boundary phenomena in simpler settings. Topolectric circuits provide such a route by mapping tight-binding models onto networks of linear circuit elements \cite{lee18,imhof18,ezawa18,hofmann19,ezawa19,liu20,yang20,helbig20,ezawa20,wu20,dong21,haydar23,haydar25,tang25}, enabling direct experimental access to topological signatures through electrical response measurements \cite{lee18}.

The topological properties of tight-binding models may be categorized according to the ten-fold way classification of nonunitary symmetries \cite{altland97, schnyder08, kitaev09, qi10, ryu10, kane10, chiu16, matveeva23}, specifically, time-reversal (TRS), charge-conjugation, and chiral symmetries \cite{altland97,shiozaki16}. Recent work has included crystalline symmetries in this classification, leading to new topological invariants and protected states \cite{teo08,fu11,hsieh12}. Crystalline symmetries may also mimic nonunitary ones, expanding the ten-fold way without requiring that the Hamiltonian satisfies further nonunitary transformations \cite{liu14,shiozaki14,young15,wang16,shiozaki16,varjas17,kruthoff17,herrera22}. A particularly important example of this in one dimension is the comparison between the symmorphic Su-Schrieffer-Heeger (SSH) model \cite{su79,su80,hasan10,asboth16,cayssol21}, with alternating hopping parameters, and the nonsymmorphic charge-density-wave (CDW) model \cite{kivelson83,shiozaki15,brzezicki20,fuchs21,cayssol21,allen22}, with constant hopping and alternating onsite energy. The SSH model is defined by its symmorphic symmetries that act locally within a unit cell, resulting in a $\mathbb{Z}$ topological index. Surprisingly, the CDW model has nonsymmorphic symmetries that involve a translation by half a unit cell, and mimic the nonunitary charge-conjugation and chiral symmetries of the SSH model \cite{mong10,fang15,shiozaki15,zhao16,yanase17,arkinstall17,otrokov19,gong19,zhang19,niu20,marques19,brzezicki20,allen22,yang22, mccann23}, resulting in a $\mathbb{Z}_2$ topological index.

\begin{figure}
    \centering
    \includegraphics[width=\linewidth]{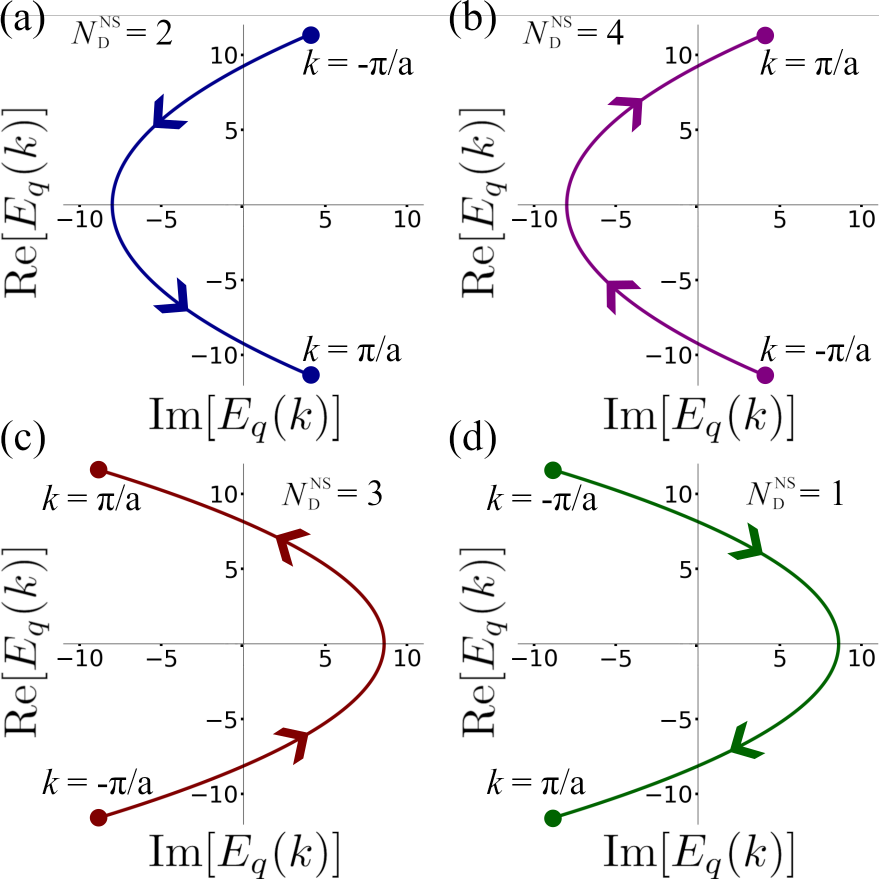}
    \caption{Four distinct topological phases of the $\mathbb{Z}_4$ model in the nonsymmorphic D class, displayed as example trajectories of the winding path $E_q(k)$, Eq.~(\ref{Eqz4}), across the Brillouin zone $-\pi/a\leq k<\pi/a$. (a) Shows a trajectory in the phase $N_\mathrm{D}^\mathrm{NS}=2$, and (b) a trajectory in the phase $N_\mathrm{D}^\mathrm{NS}=4$. Both (a) and (b) have parameter values $v=0.5$, $\Delta_s=1$, and $\phi_s=\pi/4$, they are distinguished as for (a) $\mu=-1$ and for (b) $\mu=1$. (c) Shows a trajectory in the phase $N_\mathrm{D}^\mathrm{NS}=3$, and (d) a trajectory in the phase $N_\mathrm{D}^\mathrm{NS}=1$. Both (c) and (d) have parameter values $\mu=1.7$, $v=1.2$, and $\Delta_s=0.6$, they are distinguished as for (c) $\phi_s=3\pi/4$ and for (d) $\phi_s=\pi/4$. It is possible to adiabatically deform the parameters such that the path in (a) may resemble the path in (b) without causing a phase transition. However, a change in sign of $\mu$ as displayed here must cause a transition between the two paths. A similar statement is true for a change in the sign of $\cos(\phi_s)$ between the paths in (c) and (d).}
    \label{z4paths}
\end{figure}

While example topological models for most symmetry classes in one spatial dimension can be realized with two bands \cite{mccann23}, models with Kramers degeneracy require at least four. In this paper, we build example four-band models in the nonsymmorphic AII class with a $\mathbb{Z}_2$ invariant, which has not been previously discussed, and the nonsymmorphic D class described in Ref.~\cite{tymczyszyn24} with a $\mathbb{Z}_4$ index, where the $\mathbb{Z}_4$ model has been obtained by writing the Hamiltonian in the Bogoliubov-de Gennes (BdG) representation \cite{tanaka12,alicea12,leijnse12,guo16,sato17}. To calculate both the $\mathbb{Z}_2$ and $\mathbb{Z}_4$ invariants we generalize the winding number from the BDI class, e.g. the SSH model, for nonsymmorphic chiral symmetry \cite{shiozaki15,brzezicki20}. This method has previously only been used for non-Kramers degenerate systems such as the CDW model, where the gap closes at the time-reversal invariant momenta. Here, we show that, for Kramers degenerate systems, there are topological band crossings at arbitrary $k$ values. Example paths of the generalized winding number for the $\mathbb{Z}_4$ model are shown for each of the four phases in Fig.~\ref{z4paths}.

We then extend the topolectric circuit methodology to nonsymmorphic systems, first by drawing comparison between the SSH model \cite{lee18,su79,su80} and the CDW model \cite{allen22,mccann23}, and then by describing topolectric realizations of the nonsymmorphic AII and $\mathbb{Z}_4$ chains. We find that the circuits agree with the topological theory, and allow for an experimentally measurable quantity (impedance) to be easily observed. Additionally, we further investigate the $\mathbb{Z}_4$ model due to its unique topology for noninteracting models in one dimension to understand the role of disorder in the presence of solitons between its phases. This is relevant in the context of topolectrics, as varied tolerances in circuit components effectively mimic conventional disorder effects \cite{lee18}. We find that, despite the nonsymmorphic symmetries being broken by disorder, the domain-wall bound state can remain near zero energy for particular disorder channels due to accidental emergent properties of the minimal model with nearest-neighbor hopping only.

%%%%%%%%%%%%%%%%%%%%%%%%%%%%%%%%%%%%%%%%%%%%%%%%%%%%%%%%%%%%%%%%
%%%%%%%%%%%%%%%%%%%%%%%%%%%%%%%%%%%%%%%%%%%%%%%%%%%%%%%%%%%%%%%%

In Sec.~\ref{AIInsSec} we describe a representative model of the nonsymmorphic AII class in $k$-space, and calculate its $\mathbb{Z}_2$ topology using a generalized winding number. We then use the same methodology to calculate the $\mathbb{Z}_4$ invariant in the nonsymmorphic D class in Sec.~\ref{nonsymDsec}. Topolectric realizations of nonsymmorphic systems are then discussed in Sec.~\ref{topolectricsec}, with a comparison between the SSH and CDW model in Sec.~\ref{sshcdwtopo}, followed by realizations of the AII and $\mathbb{Z}_4$ models in Sec.~\ref{aiitopo} and Sec.~\ref{z4toposec}, respectively. Finally, we add disorder in the presence of solitons to the $\mathbb{Z}_4$ model, including long-range parameters, in Sec.~\ref{disordersec}.

\section{Nonsymmorphic AII class with a $\mathbb{Z}_{2}$ index}
\label{AIInsSec}

The AII class is generally topologically trivial when only symmorphic TRS $T^{2}=-1$ is present. By introducing nonsymmorphic charge-conjugation and chiral symmetries we can induce a nontrivial $\mathbb{Z}_2$ index \cite{shiozaki16}. Note that, although charge-conjugation symmetry is present, we do not consider this model to be a superconductor as the symmetry is nonsymmorphic. The nonunitary symmetry operations can be written as
\begin{align}
    &\mathrm{time}:&U_{T}^{\dagger}(k)H^{\ast}(k)U_{T}(k)=H(-k), \\
    &\mathrm{charge}:&U_{C}^{\dagger}(k)H^{\ast}(k)U_{C}(k)=-H(-k), \\
    &\mathrm{chiral}:&U_{S}^{\dagger}(k)H(k)U_{S}(k)=-H(k),
\end{align}
where $U_{T}$, $U_{C}$, and $U_{S}$ are unitary matrices which obey the relationship $U_{S}(k)=U_{C}^{\ast}(k)U_{T}(-k)$. In $k$-space, symmorphic and nonsymmorphic symmetry operators are distinguished by the fact that the nonsymmorphic operators acquire a $k$ dependence for zero intracell spacing \cite{allen22,mccann23}. In position space the difference is clearer, as nonsymmorphic operators contain a translation of half the unit cell, as shown in Appendix~\ref{posmodels}.

\begin{figure}
    \centering
    \includegraphics[width=\linewidth]{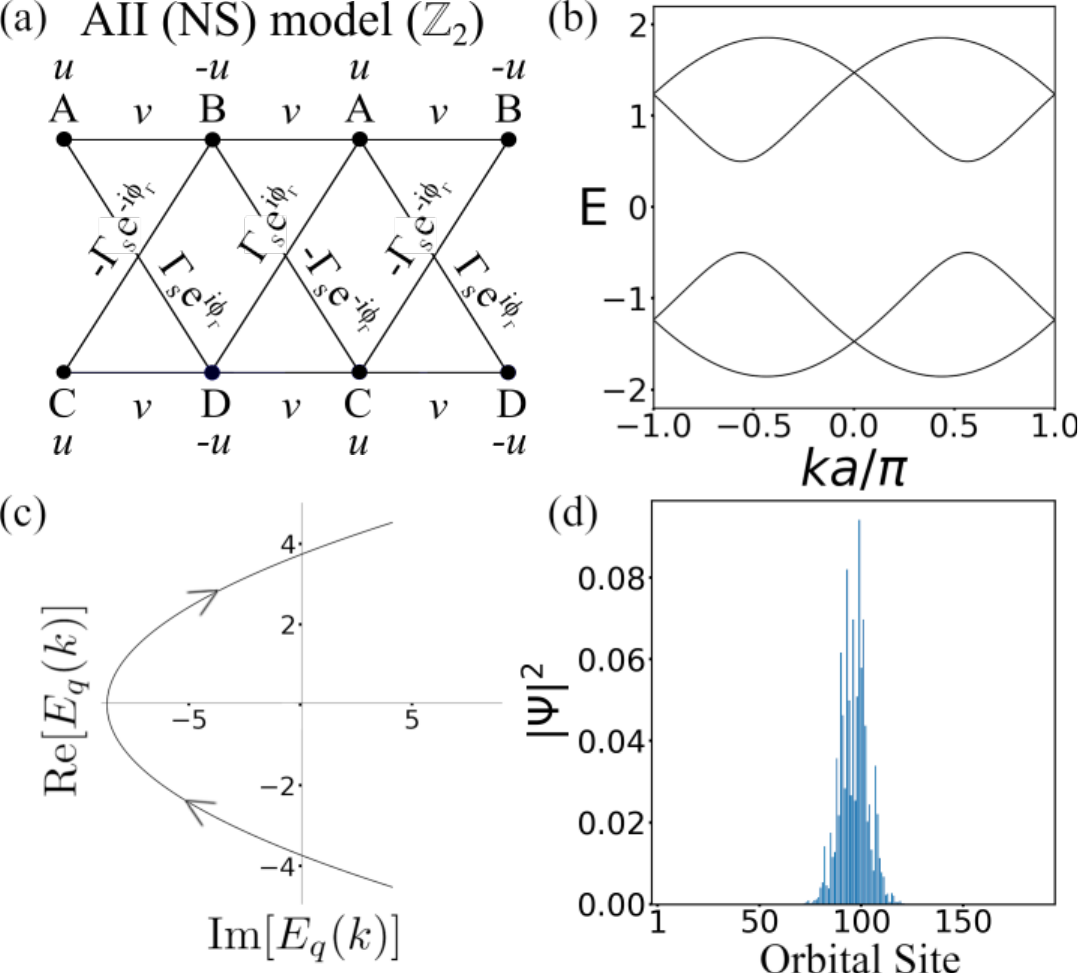}
    \caption{Nonsymmorphic AII class with symmorphic time-reversal symmetry and nonsymmorphic charge-conjugation and chiral symmetries. (a) A schematic of an example system described by Hamiltonian~(\ref{HAII}) with alternating onsite energy $\pm u$, constant nearest-neighbor hopping $v$, and constant coupling $\Gamma$ with phase $\phi_\Gamma$. (b) An example band structure, as $T^2=-1$ Kramers degeneracy is present at both $k=0$ and $k=\pi/a$, (c) shows the corresponding path of $E_q(k)$ for $N_{\mathrm{AII}}^\mathrm{NS}=1$. (d) An example soliton state for an atomically smooth texture in the onsite energy localized on the transition $u=0$ with $u_\mathrm{max}=1$, $\zeta=12$ and for 48 unit cells, with weak atomically sharp disorder in the hopping parameter $v$ of strength $W=10^{-10}$ that lifts the Kramers degeneracy. Figures (b), (c), and (d) have parameter values $v=0.4$, $\Gamma=0.8$, and $\phi_\Gamma=\pi/4$, and (b) and (c) also have a fixed value of $u=0.5$.}
    \label{AIIfig}
\end{figure}

To build the AII model, we first write a generic Hamiltonian with four atomic sites in the basis $\Psi_k^\dagger=(c^\dagger_{A,k},c^\dagger_{B,k},c^\dagger_{C,k},c^\dagger_{D,k})$. This model represents two coupled CDW chains that are time-reversal partners of each other such that the time-reversal is symmorphic with $U_T=\tau_y\sigma_0$. Atomic sites A and C are at the same spatial position, sites B and D are also at the same spatial position and are separated from A and C by half the lattice constant. The chains are described by a constant hopping $v$, alternating onsite energy $\pm u$, and an inter-chain coupling parameter $\Gamma$ with phase $\phi_\Gamma$, a schematic of the system is given in Fig.~\ref{AIIfig}(a). The bulk Hamiltonian for this system can be written as
\begin{equation}
    H(k)=
    \begin{pmatrix}
        h(k) & d(k) \\
        d^{\dagger}(k) & h^{\ast}(-k) \\
    \end{pmatrix},
    \label{HAII}
\end{equation}
where $d(k)=-d^{T}(-k)$. Imposing nonsymmorphic charge-conjugation and chiral symmetries satisfied by $U_{C}=\tau_{z}\sigma_{y}$ and $U_{S}=\tau_{x}\sigma_{y}$, we have
\begin{eqnarray*}
    h(k)&=&\!
    \begin{pmatrix}
        u & 2v\cos(ka/2) \\
        2v\cos(ka/2) & -u\\
    \end{pmatrix},\\
    d(k)&=&\!
    \begin{pmatrix}
        0 & \!2\Gamma\!\cos(ka/2\!+\!\phi_\Gamma) \\
        -\!2\Gamma\!\cos(ka/2\!-\!\phi_\Gamma) & 0 \\
    \end{pmatrix}.
\end{eqnarray*}
In general the band structure for this Hamiltonian is insulating, an example band structure is given in Fig.~\ref{AIIfig}(b), which shows Kramers degeneracy at the time-reversal invariant points $k=0$ and $k=\pi/a$. The position space representation of this model is given in Appendix~\ref{posmodels}. The topology of this model is described by a $\mathbb{Z}_{2}$ index which we can calculate by utilizing the chiral symmetry to define a $Q$-matrix, with $Q(k)=U^\dagger_{xy}H(k)U_{xy}$, where
\begin{equation*}
    Q(k) = \begin{pmatrix}
        0 & q(k) \\
        q^\dagger(k) & 0 
    \end{pmatrix},\qquad
    U_{xy}=\frac{1}{\sqrt{2}}\begin{pmatrix}
        i & 0 & -i & 0 \\
        0 & -i & 0 & i \\
        0 & 1 & 0 & 1 \\
        1 & 0 & 1 & 0
    \end{pmatrix},
\end{equation*}
such that
\begin{equation*}
    q(k)\!=\!
    \begin{pmatrix}
        2i\Gamma\cos(ka/2+\phi_\Gamma)\!-\!u & 2v\cos(ka/2) \\
        2v\cos(ka/2) & u\!+\!2i\Gamma\cos(ka/2-\phi_\Gamma) \\
    \end{pmatrix}.
\end{equation*}

%%%%%%%%%%%%%%%%%%%%%%%%%%%%%%%%%%%%%%%%%%%%%%
%%%%%%%%%%%%%%%%%%%%%%%%%%%%%%%%%%%%%%%%%%%%%%

\noindent We plot the path in the complex plane of the determinant of $q(k)$, which we denote as $E_{q}(k)$, and, for this system, we find that
\begin{eqnarray}
    E_{q}(k)&=& 4\Gamma^2\sin^2(ka/2)\!-\!4\Gamma^2\cos^2(\phi_\Gamma)\!-\!4v^2\cos^2(ka/2)\nonumber\\
    &&-u^{2}-4iu\Gamma \sin(\phi_\Gamma)\sin(ka/2)\,.
\end{eqnarray}
Similarly to the $\mathbb{Z}_4$ model, Sec.~\ref{nonsymDsec}, plotting $E_{q}(k)$ between $k=-\pi/a$ and $k=\pi/a$ results in an open path with a guaranteed crossing of the real axis at $k=0$. This is due to the $4\pi/a$ periodicity in the complex plane induced by the nonsymmorphic symmetry, and is in contrast to symmorphic chiral symmetry which produces a closed-loop winding number \cite{chiu16,ryu10,matveeva23}. An example path of $E_q(k)$ is plotted in Fig.~\ref{AIIfig}(c). As is the case of the two-band CDW model, we can attempt to find the point of the phase transition by setting $E_{q}(0)=0$, however, this yields $-u^{2}-4v^{2}-4\Gamma^2=0$. As $u^{2}$, $v^{2}$, and $\Gamma^{2}$ are all positive constants, $E_{q}(0)<0$ for all nonzero parameter values. We instead define the $\mathbb{Z}_2$ index by examining the constraints imposed on the end of the path at $k=\pi/a$. To do this we first set $\mathrm{Im}[E_q(k)]=-4u\Gamma\sin(\phi_\Gamma)\sin(ka/2)=0$, for which there are three solutions for $0<k\leq\pi/a$: $\phi_\Gamma=0$, $\Gamma=0$, and $u=0$. For these solutions $E_q(k)$ traverses only the real axis, and, as $E_q(0)<0$, the sign of $E_q(\pi/a)$ for each of these solutions determines whether the path passes through the origin, causing a phase transition. We find that $E_q(\pi/a,\phi_\Gamma=0)=E_q(\pi/a,\Gamma=0)=-u^2$, and, for nonzero $u$, these are always negative quantities such that the path of $E_q(k)$ does not pass through the origin. However, $E_q(\pi/a,u=0)=4\Gamma^2(1-\cos^2\phi_\Gamma)$ which is always positive, such that the path of $E_q(k)$ does pass through the origin, inducing a phase transition. It is not possible to adiabatically transform the parameters of $E_q(k)$ to avoid this transition, and hence there is a topological phase transition protected by the nonsymmorphic symmetry at $u=0$. As a result, the index has values given by
\begin{equation}
    N_{\mathrm{AII}}^\mathrm{NS}=\begin{cases}
        0 & \mathrm{if}\hspace{3mm}u>0\,, \\
        1 & \mathrm{if}\hspace{3mm}u<0\,.
    \end{cases}
    \label{AIIindex}
\end{equation}
We find that, similar to other nonsymmorphic models \cite{mccann23}, neither phase hosts zero-energy edge states. To emphasize the topological nature of this transition we instead consider an atomically smooth soliton in the alternating onsite energy described by the function
\begin{equation}
    u_{l}=(-1)^l\,u_{\mathrm{max}}\mathrm{tanh}\left(\frac{l-N-1/2}{\zeta}\right),
    \label{alternatingsol}
\end{equation}
for atomic site $l$, where $u_{\mathrm{max}}$ is the magnitude of the texture at infinity and $\zeta$ is the soliton width. We find that, for atomically sharp solitons ($\zeta\rightarrow0$) or smooth solitons on the length scale of the unit cell ($\zeta\gg a$), the nonsymmorphic symmetries are approximate in the presence of a soliton that breaks translational invariance. This results in energy levels that are not symmetric about $E=0$, including the soliton with finite energy $E_{\mathrm{sol}}$ that does not have a chiral partner at energy $-E_{\mathrm{sol}}$. This effect is minimized for an atomically smooth soliton described by Eq.~(\ref{alternatingsol}), the nonsymmorphic symmetries are not significantly broken as the translational symmetry local to each unit cell is mostly preserved. Therefore, the soliton energy level is pinned close to zero-energy and is doubly degenerate due to Kramers degeneracy. The zero-energy mode has a wavefunction exponentially localized onto the center of the soliton at the point $u=0$, as shown in Fig.~\ref{AIIfig}(d). To numerically obtain Fig.~\ref{AIIfig}(d) we broke the Kramers degeneracy by adding small disorder to the hopping parameter $v$, drawn randomly from a uniform distribution $[-W,W]$ of magnitude $W=10^{-10}$.

\section{Nonsymmorphic D class with a $\mathbb{Z}_{4}$ index}
\label{nonsymDsec}

\subsection{Minimal model}
\label{mintopcalc}

\begin{figure}
    \centering
    \includegraphics[width=\linewidth]{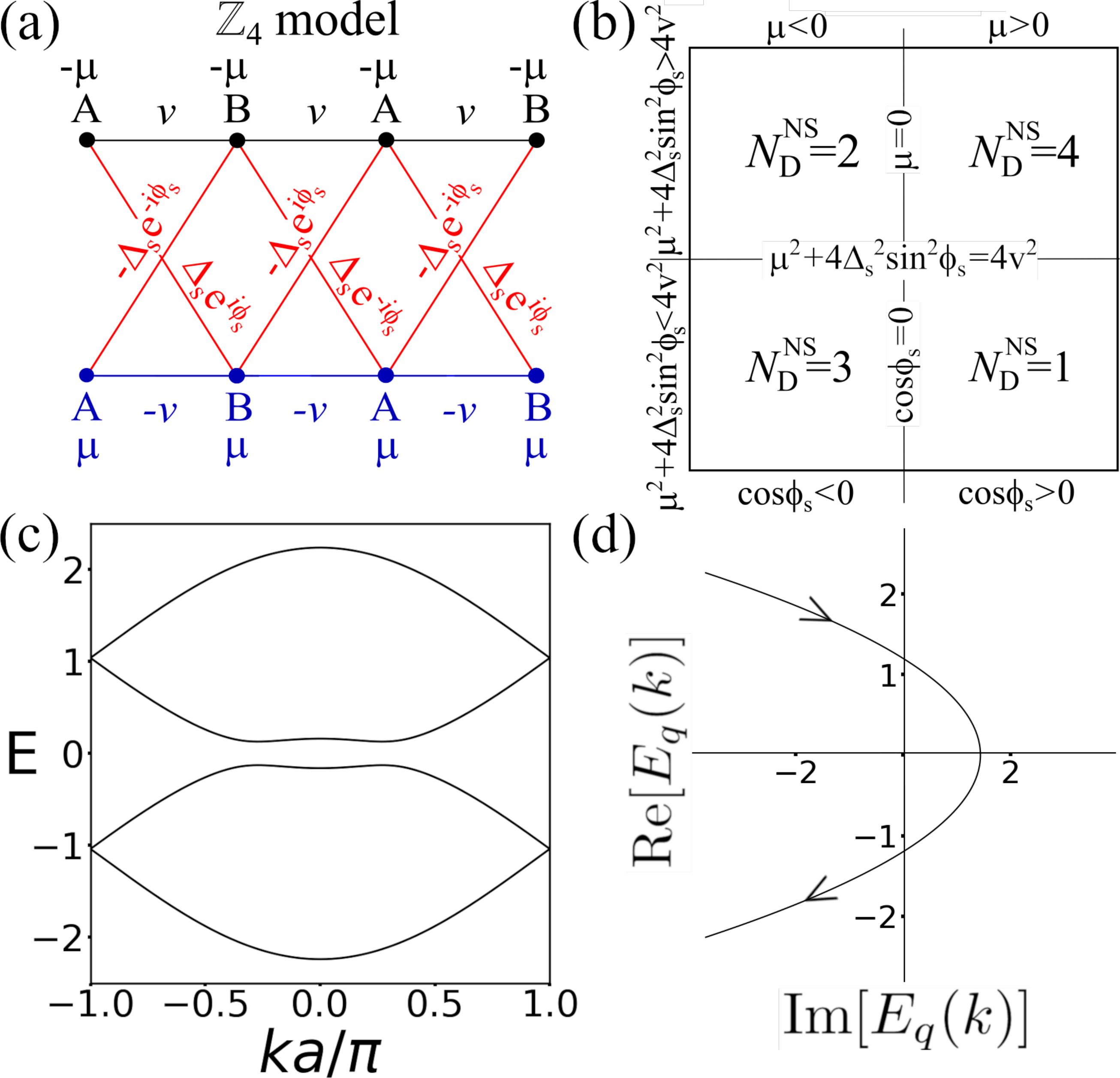}
    \caption{(a) The $\mathbb{Z}_4$ model shown in the Bogoliubov-de Gennes (BdG) representation with two orbitals A and B per cell with constant chemical potential $\mu$, nearest-neighbor hopping $v$, and magnitude of the superconducting order parameter $\Delta_s$, with alternating phase of the order parameter $\phi_s$. (b) Phase diagram of the $\mathbb{Z}_4$ topological superconductor with phases $N^{\mathrm{NS}}_{\mathrm{D}} = 1,2,3,4$. (c) Example bulk band structure $\mathrm{E}(k)$ with the corresponding path of $E_{q}(k)$ in (d) defining a phase $N_\mathrm{D}^\mathrm{NS}=3$. In (c) and (d) parameter values are $\mu=1$, $v=0.6$, $\Delta_{s}=0.2$, and $\phi_{s}=\pi/4$.}
    \label{z4fig1}
\end{figure}

The $\mathbb{Z}_2$ index belonging to the symmorphic D class, e.g., the Kitaev chain with symmorphic charge-conjugation symmetry \cite{kitaev01}, is expanded to a $\mathbb{Z}_4$ index when nonsymmorphic TRS and chiral symmetry are also enforced. We note that the $\mathbb{Z}_4$ index does not refer to the degeneracy of the ground state \cite{kitaev01,cheng12,zhang14} or unique propagation of edge currents \cite{cheong15}, but rather four distinct topological phases, with phase transitions that may be unrelated to the Majorana number. Here, we explore in more detail an example model with nonsymmorphic $U_{T}=\tau_{0}\sigma_{x}$ and $U_{S}=\tau_{x}\sigma_{x}$, as first described in Ref.~\cite{tymczyszyn24}. This model is in the BdG representation, i.e., it is represented by an electron chain coupled to a hole chain such that the symmorphic charge-conjugation operator $U_C=\tau_x\sigma_0$. The system is defined by constant chemical potential $\mu$, hopping $v$, and superconducting order parameter $\Delta_s$, and a superconducting phase $\phi_s$ that alternates sign along the length of the chain. A schematic of the model is shown in Fig.~\ref{z4fig1}(a). The alternating phase distinguishes this model from the Kitaev chain, and results in a $\mathbb{Z}_4$ index, denoted $N_\mathrm{D}^{\mathrm{NS}}$, with phases shown in Fig.~\ref{z4fig1}(b). The position space form of this model can be found in Appendix~\ref{posmodels}, and the Bloch Hamiltonian is
\begin{eqnarray}
{\ H} (k) &=& \begin{pmatrix}
\hat{h} (k) & \hat{\Delta} (k) \\
\hat{\Delta}^{\dagger} (k) & -\hat{h}^{T} (-k)
\end{pmatrix} , \label{nsham} \\
\hat{h} (k) &=& \begin{pmatrix}
- \mu & 2 v \cos (ka/2) \\
2 v \cos (ka/2)  & -\mu
\end{pmatrix} , \nonumber \\
\hat{\Delta} (k) &=& \begin{pmatrix}
0 & 2i\Delta_{s} \sin (ka/2 + \phi_{s}) \\
2i\Delta_{s} \sin (ka/2 - \phi_{s})  & 0
\end{pmatrix} . \nonumber
\end{eqnarray}

%%%%%%%%%%%%%%%%%%%%%%%%%%%%%%%%%%%%%%%%%%%%%%%%%%%%%%%%%%%
%%%%%%%%%%%%%%%%%%%%%%%%%%%%%%%%%%%%%%%%%%%%%%%%%%%%%%%%%%%

\noindent In general, the band structure for this Hamiltonian is insulating, and an example band structure is shown in Fig.~\ref{z4fig1}(c), which shows Kramer's degeneracy only at the Brillouin zone edge $k=\pi/a$ \cite{mccann23}. This is in contrast to symmorphic time-reversal symmetry, which also shows degeneracy at $k=0$. Previously, calculation of the parameter regions in which each of the $\mathbb{Z}_{4}$ phases exists was accomplished through a Pfaffian based equation \cite{shiozaki16,tymczyszyn24}. Here, we instead obtain the phases through a generalization of the winding number, as the topology can be calculated more intuitively from the utilization of the chiral symmetry, similarly to the case of the nonsymmorphic AII class described in Sec.~\ref{AIInsSec}. First, we transform Hamiltonian~(\ref{nsham}) into the form of a Q-matrix with $Q(k)=U_{xx}^{\dagger}H(k)U_{xx}$ where
\begin{equation*}
Q(k) = \begin{pmatrix}
    0 & q(k) \\
    q^\dagger (k) & 0 
\end{pmatrix},\qquad
    U_{xx}=\frac{1}{\sqrt{2}}
    \begin{pmatrix}
        -1 & 0 & 1 & 0 \\
        0 & -1 & 0 & 1 \\
        0 & 1 & 0 & 1 \\
        1 & 0 & 1 & 0 \\
    \end{pmatrix},
\end{equation*}
such that
\begin{equation*}
    q(k) \!=\! 
    \begin{pmatrix}
        \mu \!-\! 2i\Delta_{s}\!\sin(ka/2 \!+\! \phi_{s}) & -2v\cos(ka/2) \\
        -2v\cos(ka/2) & \mu\! - \!2i\Delta_{s}\!\sin(ka/2\! - \!\phi_{s})
    \end{pmatrix}.
\end{equation*}
To characterize the topology we plot the determinant of $q(k)$ in the complex plane across the Brillouin zone, i.e. for $-\pi/a\leq k <\pi/a$, where, for this system, the determinant can be written as
\begin{eqnarray}
    E_{q}(k) &=& \mu^{2}-4\Delta_s^2\sin^2(ka/2)+4\Delta_s^2\sin^2(\phi_{s}) \\
    &-&4v^2\cos^2(ka/2)-4i\mu\Delta_{s}\cos(\phi_{s})\sin(ka/2).\nonumber
    \label{Eqz4}
\end{eqnarray}
For the band structure shown in Fig.~\ref{z4fig1}(c) we can construct a corresponding path of $E_q(k)$, as shown in Fig.~\ref{z4fig1}(d). Phase transitions are defined for parameter values in which $E_{q}(k)$ passes through the origin, and that cannot be avoided by adiabatically deforming the parameters. Again, due to the nonsymmorphism, this does not produce a closed-loop winding number, but rather an open-loop path due to the $4\pi/a$ periodicity of the imaginary and real components of $E_{q}(k)$ \cite{shiozaki15}. Typically, this methodology results in a $\mathbb{Z}_{2}$ index, as the Hamiltonian symmetries enforce the path to be symmetric about the real axis such that the only topological band gap closing occurs at $E_{q}(0)$, as is the case for the CDW model. Alternatively, for Kramers degenerate systems like the aforementioned AII model, the $\mathbb{Z}_2$ index may be realized by transitions at non-zero $k$ values depending on the parameters of the system. For the $\mathbb{Z}_{4}$ model, we effectively combine the two types of transitions, i.e., transitions at $k=0$ and $k\neq 0$, to define four distinct phases.

We first examine the transition associated with $E_{q}(0)=0$. This transition is protected purely by the symmorphic charge-conjugation symmetry, and, as such, is identical to a Majorana number calculation that determines the presence of Majorana zero modes (MZM). This transition is found simply by setting $E_q(0)=0$ where
\begin{equation}
    E_{q}(0)=\mu^2+4\Delta_{s}^2\sin^2(\phi_{s})-4v^2,
    \label{Eq0Z4}
\end{equation}
such that phases are separated by whether $E_q(0)<0$ (phases with MZM) or $E_q(0)>0$ (phases without MZM). For the other two transitions we must solve $E_{q}(k)=0$ for nonzero $k$, a problem that can be simplified by considering the imaginary and real parts in isolation. Specifically, we must find parameter values such that $\mathrm{Im}[E_{q}(k)] = -4\mu\Delta_{s}\cos(\phi_{s})\sin(k/2) = 0$ for nonzero $k$. We assume that the electron and hole chains remain coupled with $\Delta_{s}\neq0$, leaving either $\mu=0$ or $\cos(\phi_{s})=0$ as solutions. For $\mu=0$ we find that $E_q(\pi)=-4\Delta_{s}^2\sin^2(\phi_{s})$, which is always less than zero. Since the path of $E_q(k)$ is confined to the real axis for $\mu=0$, this means that, for the phases with $E_{q}(0)>0$, the path must always pass through the origin, resulting in a closing of the bulk band gap that corresponds to a secondary phase transition. For the phases with $E_q(0)<0$ both ends of the path are now negative and a crossing is not guaranteed. The same logic can be applied in the case of $\cos\phi_s=0$ for which $E_q(\pi)=\mu^2$, since this is always positive, a transition is only guaranteed for $E_{q}(0)<0$. Overall, the phases are
\begin{equation*}
    N_\mathrm{D}^\mathrm{NS}\!=\!
    \begin{cases}
        1 & \mathrm{if} \hspace{2mm} \mu^{2}+4\Delta_{s}^{2}\sin^{2}(\phi_{s})<4v^{2} \hspace{2mm}\mathrm{and} \hspace{2mm}\cos\phi_{s}\!>\!0,\\
        2 & \mathrm{if}\hspace{2mm} \mu^{2}+4\Delta_{s}^{2}\sin^{2}(\phi_{s})>4v^{2} \hspace{2mm}\mathrm{and}\hspace{2mm} \mu<0, \\
        3 & \mathrm{if}\hspace{2mm} \mu^{2}+4\Delta_{s}^{2}\sin^{2}(\phi_{s})<4v^{2} \hspace{2mm}\mathrm{and}\hspace{2mm} \cos\phi_{s}\!<\!0,\\
        4 & \mathrm{if}\hspace{2mm} \mu^{2}+4\Delta_{s}^{2}\sin^{2}(\phi_{s})>4v^{2} \hspace{2mm}\mathrm{and}\hspace{2mm} \mu>0 ,
    \end{cases}
\end{equation*}
as given in the phase diagram, Fig.~\ref{z4fig1}(d). 
Therefore, we have shown that the topology of the $\mathbb{Z}_{4}$ model can be derived purely from the utilization of the chiral symmetry.

Example trajectories for each of the four phases are shown in Fig.~\ref{z4paths}. The phases $N_\mathrm{D}^{\mathrm{NS}}=2$ and $N_\mathrm{D}^{\mathrm{NS}}=4$, shown in Fig.~\ref{z4paths}(a) and Fig.~\ref{z4paths}(b), are related by a sign reversal of $\mu$. As discussed previously, this sign change necessarily drives a phase transition at $\mu = 0$. The phases $N_\mathrm{D}^{\mathrm{NS}}=3$ and $N_\mathrm{D}^{\mathrm{NS}}=1$, depicted in Fig.~\ref{z4paths}(c) and Fig.~\ref{z4paths}(d), are similarly related by a sign reversal of $\cos(\phi_s)$, implying a phase transition at $\cos(\phi_s) = 0$. It is possible, however, to adiabatically deform $\Delta_s$ and $\phi_s$ at fixed $\mu$ in Fig.~\ref{z4paths}(a) to produce a trajectory resembling Fig.~\ref{z4paths}(b) without encountering a phase transition. In contrast, changing the sign of $\mu$ invariably induces a transition and reverses the direction of the trajectory across the Brillouin zone. An analogous relationship holds between Fig.~\ref{z4paths}(c) and Fig.~\ref{z4paths}(d) when varying $\mu$ and $\Delta_s$ at fixed $\phi_s$.

\subsection{Solitons in the minimal $\mathbb{Z}_{4}$ model}

%%%%%%%%%%%%%%%%%%%%%%%%%%%%%%%%%%%%%%%%%%%%%%%%%%%%%%%%%
\begin{figure}
    \centering
    \includegraphics[width=\linewidth]{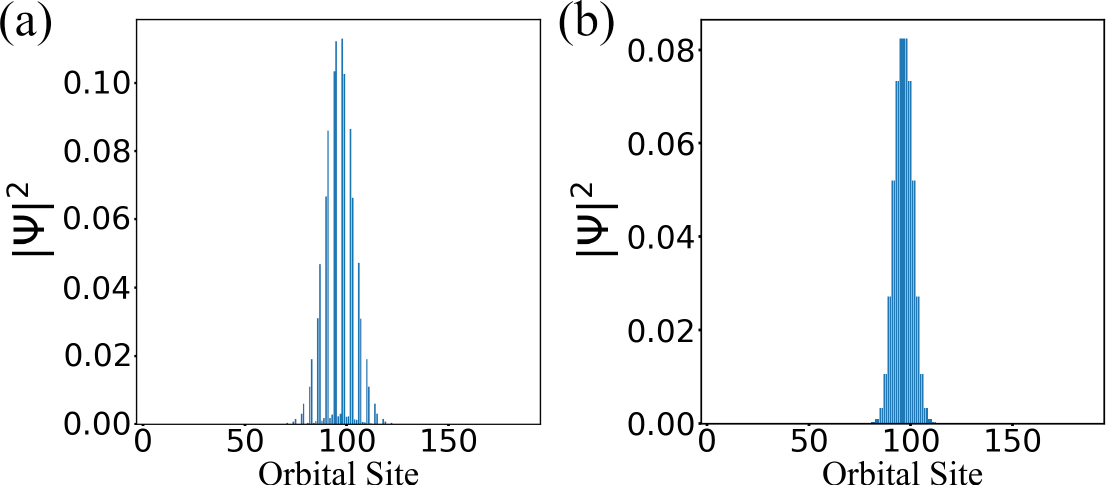}
    \caption{Probability densities of near zero-energy soliton states exponentially localized onto the center of a chain of 48 unit cells. (a) A system with a soliton in the chemical potential between phases $N_\mathrm{D}^\mathrm{NS}=2$ and $N_\mathrm{D}^\mathrm{NS}=4$, with parameter values $\Delta_{s}=1$, $v=0.3$, $\phi_s=\pi/2$, and $\mu_{\mathrm{max}}=1$. (b) A system with a soliton in the superconducting phase between phases $N_\mathrm{D}^\mathrm{NS}=1$ and $N_\mathrm{D}^\mathrm{NS}=4$, with parameter values $\Delta_{s}=1$, $v=1.5$, $\mu=2$, and $\phi_{s,\mathrm{max}}=\pi$. In both systems there exists a second state localized onto the soliton with opposite energy, for the system in (b) there also exist two Majorana zero-modes localized onto the ends of the chain.}
    \label{solwf}
\end{figure}
%%%%%%%%%%%%%%%%%%%%%%%%%%%%%%%%%%%%%%%%%%%%%%%%%%%%%%%%%

While the topological nature of the Majorana boundary at $E_q(0)=0$ is well understood \cite{kitaev01}, the additional phase transitions at $\mu=0$ and $\cos\phi_{s}=0$ are not. To provide further evidence of the topological nature for these transitions we introduce solitons into the system, where a texture in parameter values hosts a zero-energy state that is topologically protected by the symmetries of the system \cite{jackiw76,su79,su80,cayssol21,allen22}. Consider an atomically smooth soliton in the chemical potential described by the function
\begin{equation}
    \mu_{l}=-\mu_{\mathrm{max}}\mathrm{tanh}\left(\frac{l-N-1/2}{\zeta}\right)+\mu_{c}\,,
    \label{onsitesol}
\end{equation}
for atomic site $l=1,2,...,J$, where $\mu_{\mathrm{max}}$ is the magnitude of the texture at infinity for $\mu_c=0$, $N=J/2$ is the number of unit cells, and $\zeta$ is the soliton width in dimensionless units, i.e., measured in units of the lattice constant. Here, $\mu_{c}$ is a constant dependent on the other parameters of the system and is used to shift all values of the soliton across the chain. This is needed for systems with long-range parameters with transitions that do not occur at $\mu=0$, as discussed in Sec.~\ref{HOPsec}. For the minimal model we consider here, we set $\mu_{c}=0$ such that the soliton is centered on $\mu=0$.

As the solitons break translational invariance they also weaken the nonsymmorphic symmetries, however, for a large enough system size and width $\zeta$ the symmetry is approximately maintained, allowing the solitons to remain localized \cite{shiozaki15,zhao16,brzezicki20,mccann23,muten24,han20}. We find that the addition of the texture results in a zero energy state exponentially localized onto the soliton at the center of the chain where $\mu\approx0$. Therefore, the soliton marks the boundary between phases $N_\mathrm{D}^\mathrm{NS}=2$ and $N_\mathrm{D}^\mathrm{NS}=4$, Fig.~\ref{z4fig1}(b). An example soliton state highlighting the exponential localization at the center of the chain is shown in Fig.~\ref{solwf}(a), with $\mu_{\mathrm{max}}=1$, $\Delta_{s}=1$, $v=0.3$, and $\phi_{s}=\pi/2$.

We also consider the case of a smooth soliton in $\phi_{s}$ described by the function
\begin{equation}
    \phi_{s,n}=\frac{\phi_{s,\mathrm{max}}}{2}\left(1+\mathrm{tanh}\left(\frac{n-N}{\zeta}\right)\right) +\phi_{s,c}\,\,,
    \label{SPsol}
\end{equation}
where $\phi_{s,\mathrm{max}}$ is the magnitude of the texture at infinity, and $n=1,2,...,J-1$ is an index representing the position between adjacent atomic sites. For the minimal model we consider here, we set $\phi_{s,c}=0$ such that the soliton is centered on $\phi_{s}=\phi_{s,\mathrm{max}}/2$. An example probability density for $\phi_{s,\mathrm{max}}=\pi$, $\Delta_{s}=1$, $v=1.5$, and $\mu=1$ is shown in Fig.~\ref{solwf}(b). In this regime, there are two zero energy states hosted by the soliton in $\phi_s$, additionally, there are two MZM exponentially localized onto the ends of the chain, as the model exists in phases with $E(0)<0$. The soliton states are localized at the center of the chain where $\phi_{s}\approx\pi/2$, the same parameter value predicting the phase transition between phases $N_\mathrm{D}^\mathrm{NS}=1$ and $N_\mathrm{D}^\mathrm{NS}=3$.

\section{Topolectric circuit realization of nonsymmorphic models}
\label{topolectricsec}
\subsection{General setup and two-point impedance}

%%%%%%%%%%%%%%%%%%%%%%%%%%%%%%%%%%%%%%%%%%%%%%%%%%%%%%%%%%%%%%%
\begin{figure}
    \centering
    \includegraphics[width=\linewidth]{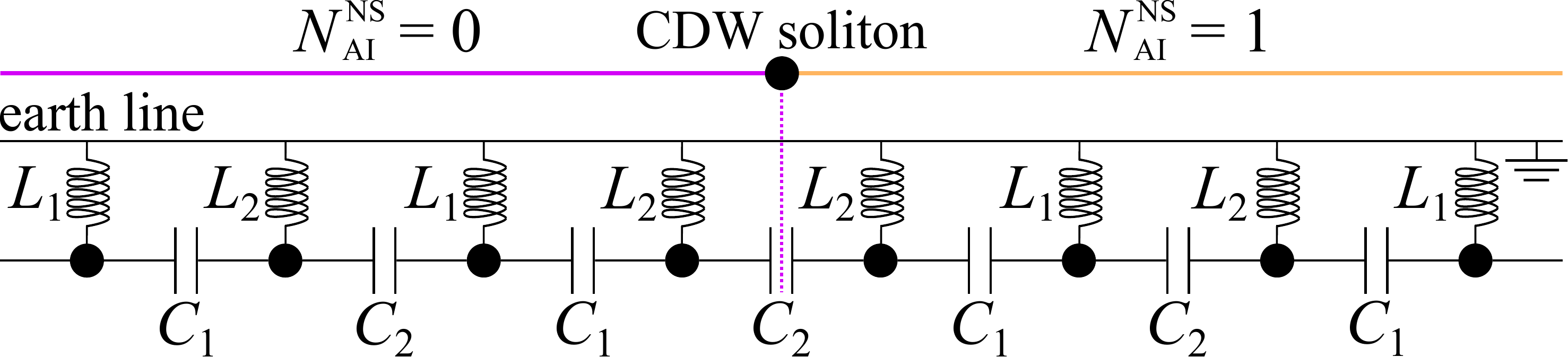}
    \caption{Schematic of a topolectric circuit realization of the SSH and CDW models. Each node is connected to adjacent nodes with staggered capacitors $C_1$ or $C_2$ and connected to the ground through alternating inductors $L_1$ or $L_2$. The SSH phase is realized for $L_2=L_1$ and the CDW phase for $C_1=C_2$. In the CDW phase an atomically sharp soliton may be implemented by swapping the order of $L_1$ and $L_2$, as shown at the center of the diagram.}
    \label{SSHCDWschematic}
\end{figure}
%%%%%%%%%%%%%%%%%%%%%%%%%%%%%%%%%%%%%%%%%%%%%%%%%%%%%%%%%%%%%%%

The use of electric circuits to realize topological systems has been studied extensively \cite{lee18,imhof18,ezawa18,hofmann19,ezawa19,ezawa20,dong21,haydar23,haydar25}; however, circuits representing nonsymmorphic models have not been explored previously. Here, we explicitly write out circuit realizations of the symmorphic SSH chain and compare it to the nonsymmorphic CDW chain, which can be realized as different phases of the same model, Fig.~\ref{SSHCDWschematic}. We then expand the nonsymmorphic methodology to topolectric realizations of the previously described nonsymmorphic AII and $\mathbb{Z}_4$ chains. First, let us define a general setup for topolectric circuits in addition to the measurable observable of impedance. The electric current that flows into the ground and the voltage of an RLC circuit at node $a$ are related by Kirchhoff's law \cite{ezawa19,ezawa20,dong21} as
\begin{eqnarray}
    \frac{d}{dt}I_m&=&\sum_n \Big[\Big(C_{mn}\frac{d^2}{dt^2}+\frac{1}{L_{mn}}+\frac{1}{R_{mn}}\frac{d}{dt}\Big)(V_m-V_n)\Big]\nonumber\\
    &&+\Big(C_m\frac{d^2}{dt^2}+\frac{1}{L_m}+\frac{1}{R_m}\frac{d}{dt}\Big)V_m\,,
    \label{difkirch}
\end{eqnarray}
where $I_m$ is the current between node $m$ and the ground, $V_m$ and $V_n$ are
the voltages at node $m$ and $n$, $C_{mn}$, $L_{mn}$, and $R_{mn}$ are the capacitance, inductance, and resistance between nodes $m$
and $n$, $C_m$, $L_{m}$, and $R_m$ are the capacitance, inductance and resistance at node $m$, and the
sum is taken over all adjacent nodes $n$. We note that, by convention, we take $C_{mn}$, $L_{mn}$, $R_{mn}$, $C_m$, $L_m$, and $R_m$ to be positive. By performing a Fourier transform $V(t)=V_0 e^{i\omega t}$, we can rewrite Eq.~(\ref{difkirch}) as\begin{equation}
    I_m(\omega)=\sum_n J_{mn}(\omega)V_n(\omega),
\end{equation}
where
\begin{widetext}
\begin{eqnarray}
    J_{mn}(\omega)=i\omega\Bigg[\!-\!C_{mn}+\!\frac{1}{\omega^2L_{mn}}\!+\!\frac{i}{\omega R_{mn}} \!+\!\delta_{mn}\bigg(\!C_m\!-\!\frac{1}{\omega^2L_m}-\frac{i}{\omega R_m}\!+\!\sum_c\Big[C_{mc}\!-\!\frac{1}{\omega^2L_{mc}}\!-\!\frac{i}{\omega R_{mc}}\Big]\bigg)\Bigg],
    \label{genlaplacian}
\end{eqnarray}
\end{widetext}
where $J_{mn}(\omega)$ is the circuit Laplacian. If the circuit Laplacian is at its resonant frequency, with respect to the intended model, then it can be related to the position-space tight-binding Hamiltonian $\mathcal{H}$ via the relation $J_{mn}(\omega)=i\omega \mathcal{H}_{mn}(\omega)$, or, in $k$-space, as $J_{mn}(k)=i\omega H_{mn}(k)$. The topological equivalence of the circuit to the tight-binding Hamiltonian is confirmed by the experimentally measurable quantity of impedance under an applied current \cite{dong21}. We inject an incoming current $I_m(\omega)$ into node $m$, resulting in an outgoing current $I_n(\omega)=I_m(\omega)$ at node $n$. The voltage difference between these two nodes then defines the two-point impedance as
\begin{equation}
    Z_{mn}(\omega)=\frac{V_m(\omega)-V_n(\omega)}{I(\omega)}=\sum_p\frac{\vert\psi_{p,m}-\psi_{p,n}\vert}{j_p}\,,
    \label{2PI}
\end{equation}
where we have diagonalized the circuit Laplacian to obtain eigenvalues $j_p$ and corresponding eigenmodes with components $\psi_{p,m}$ at node $m$. When at least one eigenvalue $j_p$ is small, $Z_{mn}$ diverges provided that $\psi_{p,m}\neq\psi_{p,n}$. Therefore, edge states, MZM, and solitons may be detected by measuring the two-point impedance.

We note that, throughout this section, we use component values that may be physically unrealizable experimentally, e.g., unrealistically large capacitances. This choice is only for numerical simplicity, and values may be scaled where appropriate. By scaling all linear components (for fixed resistance) we identically scale the eigenvalues of the circuit Laplacian, which inversely scales the impedance magnitude, Eq.~(\ref{2PI}). In general, this will only effect the absolute magnitude of the impedance peaks, with the underlying physics remaining unaffected.

\subsection{Circuit realizations of the SSH and CDW models}
\label{sshcdwtopo}

%%%%%%%%%%%%%%%%%%%%%%%%%%%%%%%%%%%%%%%%%%%%%%%%%%%%%%%%%%%%%%%
\begin{figure}
    \centering
    \includegraphics[width=\linewidth]{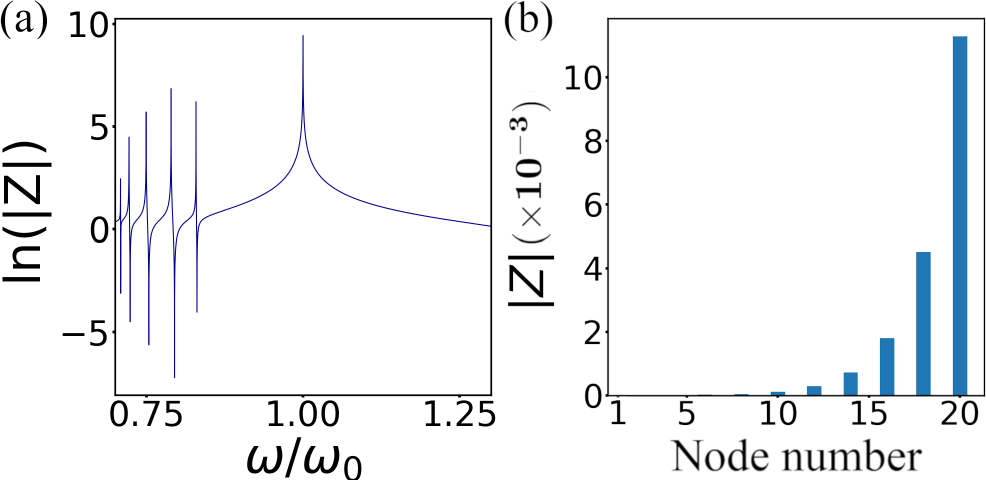}
    \caption{Two-point impedance magnitudes across topolectric realizations of the SSH model with 20 nodes in the topological phase $C_1/C_2=0.4$ with edge states and current input at the first site in the chain. In (a) the current output is at the edge state on the opposite end of the chain and the logarithm of the magnitude of the impedance $\ln(|\mathrm{Z|)}$ is plotted as a function of the frequency, while (b) shows the on-resonance impedance as a function of variable output node, where the value at site 20 corresponds to the peak in (a) at $\omega/\omega_0=1$.}
    \label{impedencegraphsSSH}
\end{figure}
%%%%%%%%%%%%%%%%%%%%%%%%%%%%%%%%%%%%%%%%%%%%%%%%%%%%%%%%%%%%%%%

The SSH and CDW models have circuit realizations that can be considered as different phases of one another. To see this, consider a chain of nodes connected by staggered capacitors $C_1$ and $C_2$ and grounded by inductors $L_1$ and $L_2$ that alternate along the chain at nodes A and B, respectively. With only these components, Eq.~(\ref{genlaplacian}) reduces to the position space Hamiltonian
\begin{eqnarray}
    J_{mn}(\omega)=i\omega\!\left[-C_{mn}\!+\!\delta_{mn}\bigg(\!\!-\frac{1}{\omega^2 L_m}+\sum_c C_{mc}\!\bigg)\!\right],
    \label{CDWSSHham}
\end{eqnarray}
where $C_{mn}=C_1$ and $L_m=L_1$ for odd values of $m$ and $C_{mn}=C_2$ and $L_m=L_2$ for even values of $m$. The SSH phase is realized when $L_2=L_1$, such that, for an infinite circuit with intracell spacing $s=0$, this results in
\begin{equation}
\renewcommand\arraystretch{1.4}
    J_{\mathrm{SSH}}(k)=i\omega\!\begin{pmatrix}
        \!C_1\!+\!C_2\!-\!\frac{1}{\omega^2L_1} & -C_1\!-\!C_2e^{-ika} \\
        -C_1\!-\!C_2e^{ika} & \!C_1\!+\!C_2\!-\!\frac{1}{\omega^2L_1}\!
    \end{pmatrix},
\end{equation}
where at the resonant frequency $\omega_0=1/\sqrt{L_1(C_1+C_2)}$, the onsite terms are zero and we find exact agreement with the SSH model with $C_1=-v$ and $C_2=-w$. Therefore, for $C_1/C_2<1$, the model is in the topological regime and hosts edge states at zero energy on the ends of the chain, while, for $C_1/C_2>1$, the model is in the trivial phase. We note that, as $C_1$ and $C_2$ are both positive, we cannot realize regimes with $C_1/C_2<0$, although this is possible by introducing subnodes into the system \cite{dong21}. By varying the frequency $\omega$ across a system in the topologically nontrivial regime ($C_1/C_2<1$), we find that there is a distinct peak in the two-point impedance between the two ends of the chain, where $l$ is the last node in the chain, as predicted by Eq.~(\ref{2PI}) and shown in Fig.~\ref{impedencegraphsSSH}(a). This peak corresponds to the zero-energy edge state of the SSH model. In Fig.~\ref{impedencegraphsSSH}(b), we plot the two-point impedance for each node across the chain when the system is at the resonant frequency and for an input current injected at the first node. There is a distinct peak in the impedance at the end of the chain that matches the peak found in Fig.~\ref{impedencegraphsSSH}(a). Due to the underlying sublattice symmetry of the SSH model, the impedance is near-zero if the current output is on an A node and non-zero on B nodes \cite{dong21}.

%%%%%%%%%%%%%%%%%%%%%%%%%%%%%%%%%%%%%%%%%%%%%%%%%%%%%%%%%%%
\begin{figure}
    \centering
    \includegraphics[width=\linewidth]{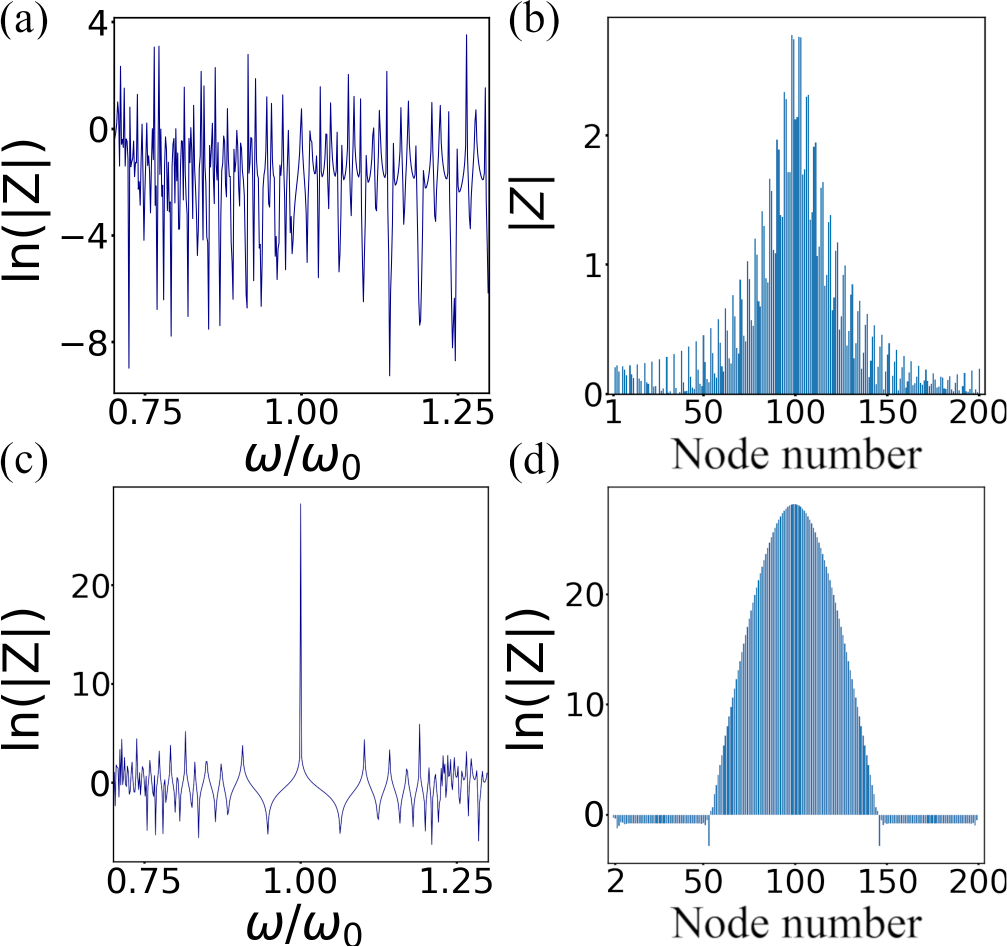}
    \caption{Two-point impedance magnitudes across topolectric realizations of the CDW model with 200 nodes in the presence of a topological soliton between the two phases and with current input at the first site in the chain. (a) and (b) are for an atomically sharp soliton, Fig.~\ref{SSHCDWschematic}, located at the center of the chain, (a) is the logarithm of the magnitude of the impedance $\ln(|\mathrm{Z|)}$ plotted as a function of the frequency with current output at the position of the soliton at the center of the chain. (b) is a plot of the on-resonance impedance as a function of output node, where the peak at the center of the chain corresponds to the peak in (a) at $\omega/\omega_0=1$. (c) and (d) are for the same system as (a) and (b), but in the presence of an atomically smooth soliton of width $\zeta=25$. We find a significantly larger peak in the impedance for a smooth soliton when compared to the sharp soliton in (a) and (b). Figures (a) and (b) use parameter values $C_1=10$F, $L_1=0.2$H, and $L_2=0.21$H and (c) and (d) use $C_1=1$F, $u_\mathrm{max}=1$.}
    \label{impedancegraphsCDW}
\end{figure}
%%%%%%%%%%%%%%%%%%%%%%%%%%%%%%%%%%%%%%%%%%%%%%%%%%%%%%%%%%%

The CDW phase is realized by setting $C_2=C_1$ in Eq.~(\ref{CDWSSHham}). For an infinite circuit, with each A and B site equally separated by half the lattice constant, this results in
\begin{equation}
\renewcommand\arraystretch{1.4}
    J_{\mathrm{CDW}}(k)=i\omega\begin{pmatrix}
        2C_1-\frac{1}{\omega^2L_1} & C_1\cos(ka/2) \\
        C_1\cos(ka/2) & 2C_1-\frac{1}{\omega^2L_2}
    \end{pmatrix}.
\end{equation}
The CDW model is characterized by an alternating onsite energy $\pm u$ which we obtain by setting $u=2C_1-1/\omega^2L_1=-2C_1+1/\omega^2L_2$, a condition satisfied by the resonant frequency
\begin{align}
    \omega_0=\frac{1}{2}\sqrt{\frac{1}{C_1L_1}+\frac{1}{C_1L_2}}\,\,.
    \label{cdwresonant}
\end{align}
The constant hopping of the CDW model is related to the capacitance as $v=-C_1$. At the resonant frequency, the two topological phases are then defined by whether $2C_1-1/\omega^2L_1>0$ $(N_\mathrm{AI}^\mathrm{NS}=0)$ or $2C_1-1/\omega^2L_1<0$  $(N_\mathrm{AI}^\mathrm{NS}=1)$. While the CDW model does not have zero energy edge states, it can host a zero energy state exponentially localized onto a soliton between the two phases \cite{allen22}. An atomically sharp soliton can be implemented by swapping the values of $L_1$ and $L_2$ at the center of the chain. The two point impedance for current injected at the edge of the chain and with current output at the center of the chain is shown as a function of the frequency $\omega$ in Fig.~\ref{impedancegraphsCDW}(a). A sharp peak in the impedance is found at the resonant frequency, and we find that the magnitude of the peak is small when compared to the SSH model, Fig.~\ref{impedencegraphsSSH}, due to the nonsymmorphic symmetry only being approximate in the presence of a soliton, requiring large system sizes to be close to zero energy \cite{allen22}. We also plot the two-point impedance for current injected at the edge of the chain for each node in the chain in Fig.~\ref{impedancegraphsCDW}(b). This shows a distinct peak in the impedance at the position of the soliton at the center of the chain.

Although more difficult to implement experimentally, requiring careful tuning of circuit components, a larger peak in the impedance may be realized by considering an atomically smooth soliton in the onsite terms. By fixing the frequency as $\omega=\omega_0$, we can define a smooth soliton by continuously changing the value of $L_m$ according to 
\begin{equation}
    L_m\!=\left[\omega_0^2\!\left(2C_1\!-\!(-1)^mu_\mathrm{max}\!\tanh\!\!\left(\!\frac{m\!-\!(N\!+\!1)/2}{\zeta}\!\right)\right)\right]^{-1}\!,
    \label{indsol}
\end{equation}
where $u_\mathrm{max}$ is the value of the alternating onsite energy at infinity, $N$ is the number of nodes, and $\zeta$ is the soliton width. This results in a significantly larger peak in the impedance as there are eigenvalues $j_p$ closer to zero energy when compared to an atomically sharp soliton \cite{allen22}. We show this in Fig.~\ref{impedancegraphsCDW}(c) as a function of the frequency for current injected at the first node and with output current at the center of the chain, where a significantly larger peak at $\omega=\omega_0$ is shown as compared to Fig.~\ref{impedancegraphsCDW}(a). In Fig.~\ref{impedancegraphsCDW}(d), the two-point impedance for current injected at the first node as a function of output node is shown, where the impedance is exponentially (linearly in the log scale) localized around the location of the soliton at the center of the chain. For both the SSH and CDW models, the component values may be scaled linearly with the magnitude of impedance spectra shown in Fig.~\ref{impedencegraphsSSH} and Fig.~\ref{impedancegraphsCDW}. 

We note that it is also possible to build a CDW circuit by alternating the grounding component from capacitors to inductors, however, this cannot realize a phase of the SSH model.

\subsection{Circuit realization of the nonsymmorphic AII model}
\label{aiitopo}

%%%%%%%%%%%%%%%%%%%%%%%%%%%%%%%%%%%%%%%%%%%%%%%%%%%%%%%%%
\begin{figure}
    \centering
    \includegraphics[width=\linewidth]{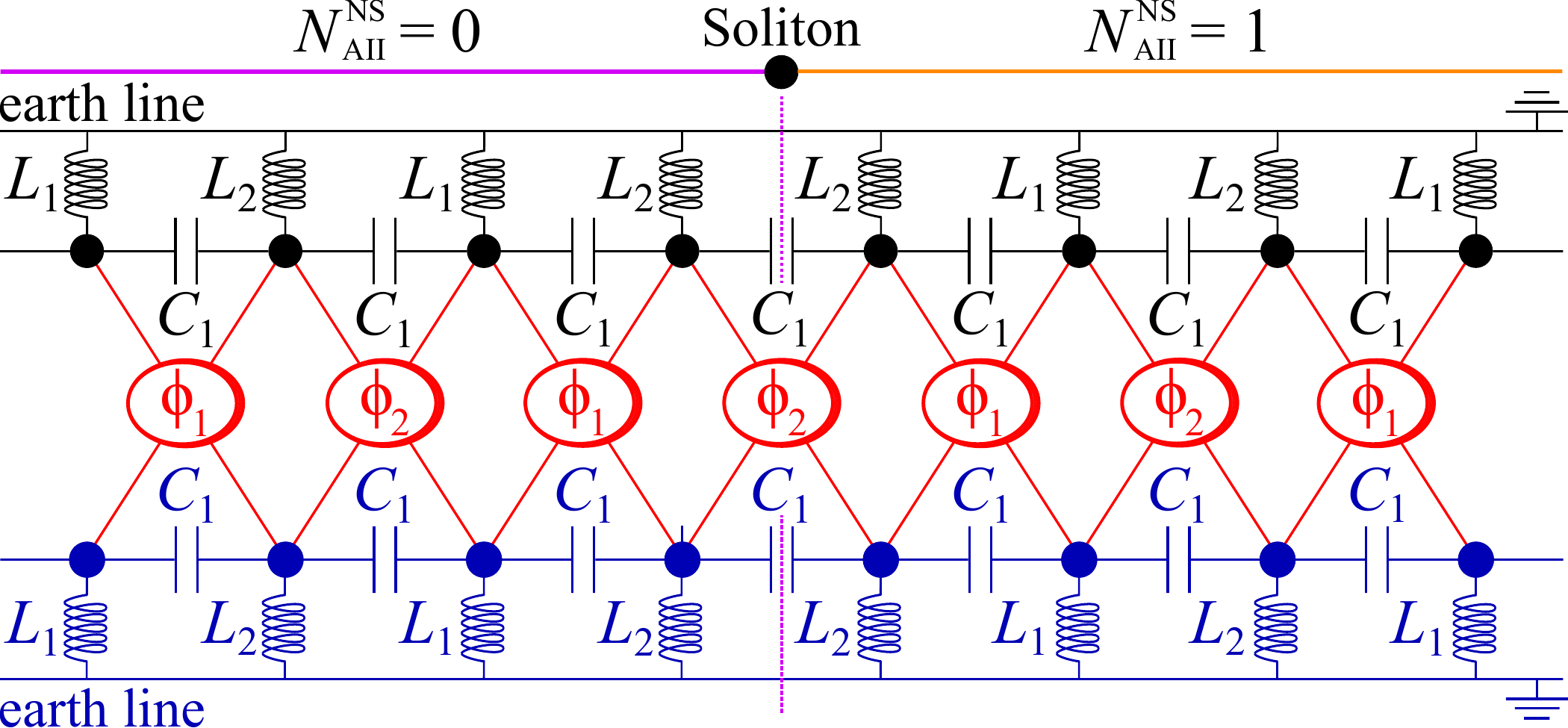}
    \caption{Topolectric circuit realization of the AII model, a soliton causes a phase transition between phases $N_\mathrm{AII}^\mathrm{S}=0$ and $N_\mathrm{AII}^\mathrm{S}=1$. The model is simulated by two channels, the upper channel (black) and lower channel (blue) are time-reversal partners of one another. The chains are grounded by either inductors $L_1$ or $L_2$, nodes in the same channel are connected by capacitors $C_1$, and the channels are paired to one another (red) by phase-control units, Fig.~\ref{PCUs} \cite{ezawa19,ezawa20}.}
    \label{topoaIIschematic}
\end{figure}
%%%%%%%%%%%%%%%%%%%%%%%%%%%%%%%%%%%%%%%%%%%%%%%%%%%%%%%%%

%%%%%%%%%%%%%%%%%%%%%%%%%%%%%%%%%%%%%%%%%%%%%%%%%%%%%%%%%
\begin{figure}
    \centering
    \includegraphics[width=0.85\linewidth]{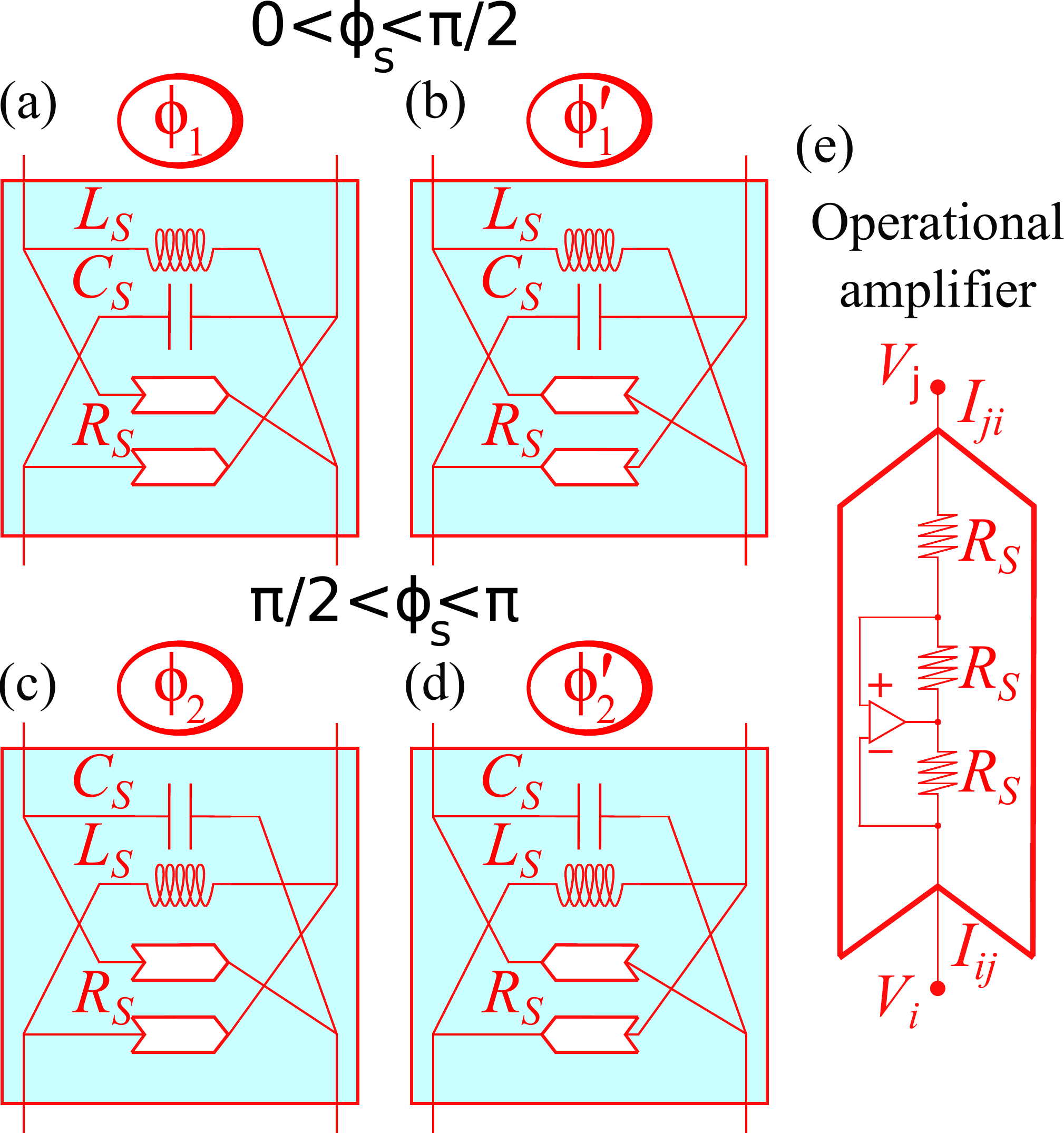}
    \caption{Circuit diagrams of the phase-control units (PCUs). Four forms of PCU are shown. The combination of (a) and (c) shows the intracell and intercell PCUs used in the AII model shown in Fig.~\ref{topoaIIschematic}. The combination of (a) and (b) shows the intracell and intercell PCUs used in the $\mathbb{Z}_4$ model shown in Fig.~\ref{topoz4schematic}, mimicking the superconducting pairing $\Delta_s$ with phase $\phi_s$ for $0<\phi_s<\pi/2$. The combination of (c) and (d) implements the same parameters for $\pi/2<\phi_s<\pi$. Figure (e) shows the structure of the operational amplifier used in (a)-(d).}
    \label{PCUs}
\end{figure}
%%%%%%%%%%%%%%%%%%%%%%%%%%%%%%%%%%%%%%%%%%%%%%%%%%%%%%%%%

%%%%%%%%%%%%%%%%%%%%%%%%%%%%%%%%%%%%%%%%%%%%%%%%%%%%%%%%%
\begin{figure}
    \centering
    \includegraphics[width=\linewidth]{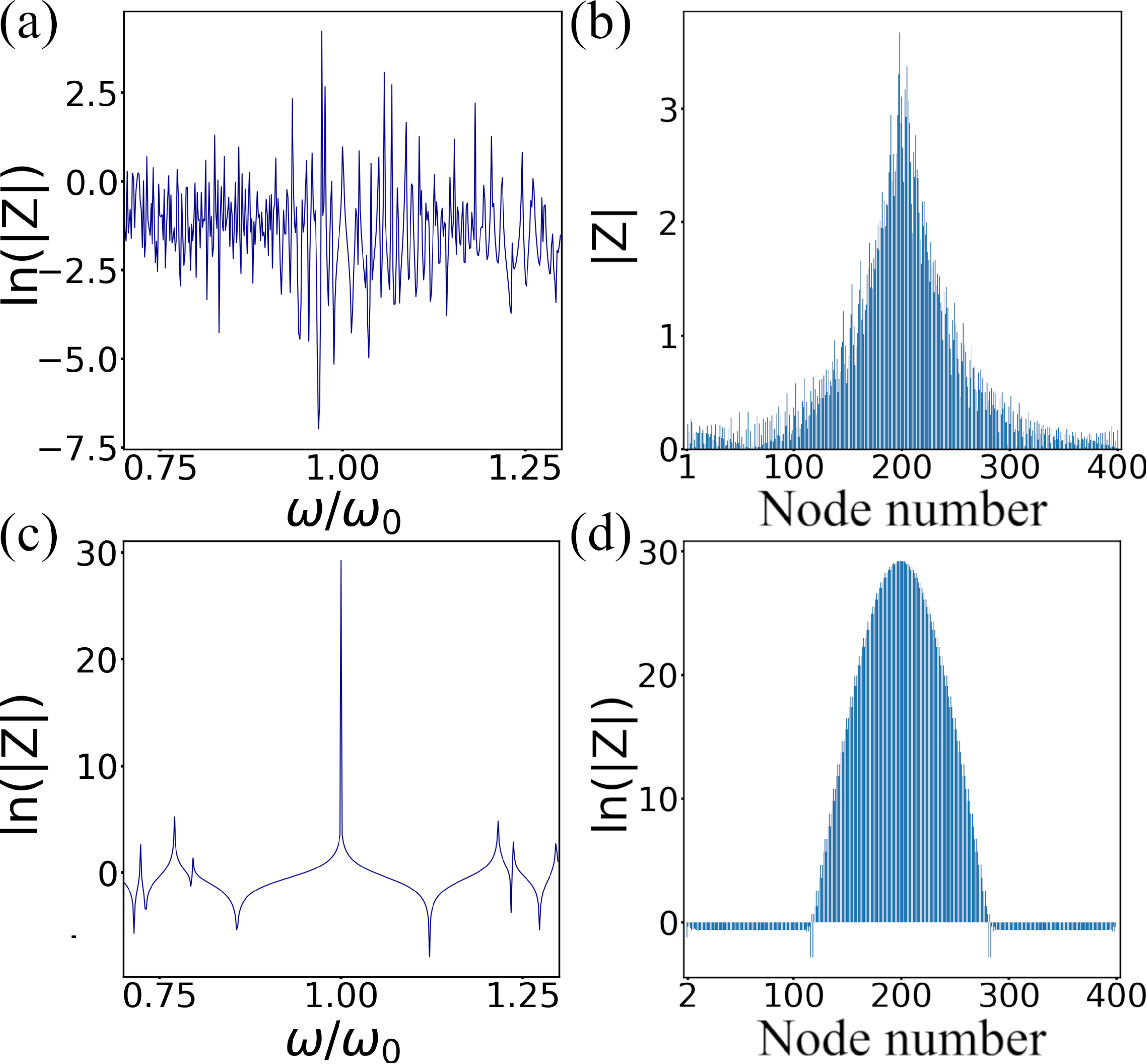}
    \caption{Two-point impedance magnitudes across topolectric realizations of the nonsymmorphic AII model, Fig.~\ref{topoaIIschematic}, for 400 nodes and input current at the first node in the upper channel. Figures (a) and (b) are for an atomically sharp soliton located at the center of the chain, (a) is the logarithm of the magnitude of the impedance $\ln(|\mathrm{Z|)}$ plotted as a function of the frequency with current output at the position of the soliton at the center of the chain. (b) is a plot of the on-resonance impedance as a function of output node, where the peak at the center of the chain corresponds to the peak in (a) at $\omega/\omega_0=1$. (c) and (d) are for the same system as (a) and (b), but in the presence of an atomically smooth soliton of width $\zeta=25$. We find a significantly larger peak in the impedance for a smooth soliton when compared to the sharp soliton in (a) and (b). Figures (a) and (b) use parameter values $C_1=4$F, $L_1=0.128$H, $L_2=0.129$H, $C_S=0.5$, $L_S=2$, and $R_S=2$, while (c) and (d) use $C_1=0.4$F, $u_\mathrm{max}=1$, $C_S=0.5$, $L_S=2$, and $R_S=2$.}
    \label{AIItopo}
\end{figure}
%%%%%%%%%%%%%%%%%%%%%%%%%%%%%%%%%%%%%%%%%%%%%%%%%%%%%%%%%

The AII chain, Eq.~(\ref{HAII}), can be modeled with two coupled CDW channels that are time-reversal partners of each other, as shown in Fig.~\ref{topoaIIschematic} \cite{ezawa19,ezawa20}. This is of a similar form to the circuit description of the Kitaev chain found in Ref.~\cite{ezawa20}. A soliton at the center of the chain marks the boundary between topological phases $N_\mathrm{AII}^\mathrm{NS}=0$ and $N_\mathrm{AII}^\mathrm{NS}=1$, Eq.~(\ref{AIIindex}). The channels are grounded by inductors $L_1$ or $L_2$, nodes are connected by capacitors $C_1$, and the two channels are coupled by the phase control units shown in Fig.~\ref{PCUs}, which are themselves composed of capacitors $C_S$, inductors $L_S$, and operational amplifiers $R_S$. The operational amplifiers, while experimentally complex, are required as they act as directional resistors, allowing for the addition of complex phases without breaking Hermiticity, i.e., flipping the current direction flips the sign of $R_S$. The resonant frequency of this system requires high specificity of the circuit components, as it is identical to the CDW resonant frequency, Eq.~(\ref{cdwresonant}), with further constraints. We find that
\begin{equation}
    \omega_0 = \frac{1}{2}\sqrt{\frac{1}{C_1L_1}+\frac{1}{C_1L_2}} = 1/\sqrt{C_SL_S}\,.
\end{equation}
The bulk circuit Laplacian can be derived by Fourier transforming the position space circuit Laplacian, Eq.~(\ref{genlaplacian}), such that, at the resonant frequency, we find that the parameters of the AII Hamiltonian~(\ref{HAII}) can be parameterized in terms of the circuit components as
\begin{eqnarray}
    u &=& 2C_1- 4C_1L_2/(L_2+L_1),\\
    v &=& -C_1, \\
    \Gamma &=& \sqrt{C_S^2+L_SC_S/R_S^2} ,\\
    \phi_\Gamma &=&\mathrm{arg}\,\left[ -C_S + i\sqrt{L_SC_S}/R_S \right]\,.
\end{eqnarray}
From this comparison we can create atomically sharp solitons at the resonant frequency by swapping $L_1$ and $L_2$, as we did previously in the CDW model, which shares the same transition point at $u=0$. Fig.~\ref{AIItopo}(a) shows the impedance as a function of input frequency, with current output at the first node in the upper chain and the output at the center of the upper chain, while Fig.~\ref{AIItopo}(b) shows the two-point on-resonance impedance for variable output node. Similarly to the CDW model, the atomically sharp soliton does not produce an impedance peak at the resonant frequency that is discernible from the background noise, although the small peak produced does appear localized at the center of the chain indicating a phase boundary. In contrast to this, for a smooth soliton described by Eq.~(\ref{indsol}), where both channels maintain identical component values to each other, the impedance peak is significantly more pronounced. The impedance as a function of input frequency for current input at the first node and output at the center of the chain is shown in Fig.~\ref{AIItopo}(c), which shows a significantly larger peak at the resonant frequency when compared to the atomically sharp soliton. The on-resonance impedance as a function of output node is shown in Fig.~\ref{AIItopo}(d), which shows exponential (linear in the log scale) localization around the location of the soliton at the center of the chain.

\subsection{Circuit realizations of the $\mathbb{Z}_4$ model}
\label{z4toposec}

%%%%%%%%%%%%%%%%%%%%%%%%%%%%%%%%%%%%%%%%%%%%%%%%%%%%%%%%%
\begin{figure}
    \centering
    \includegraphics[width=\linewidth]{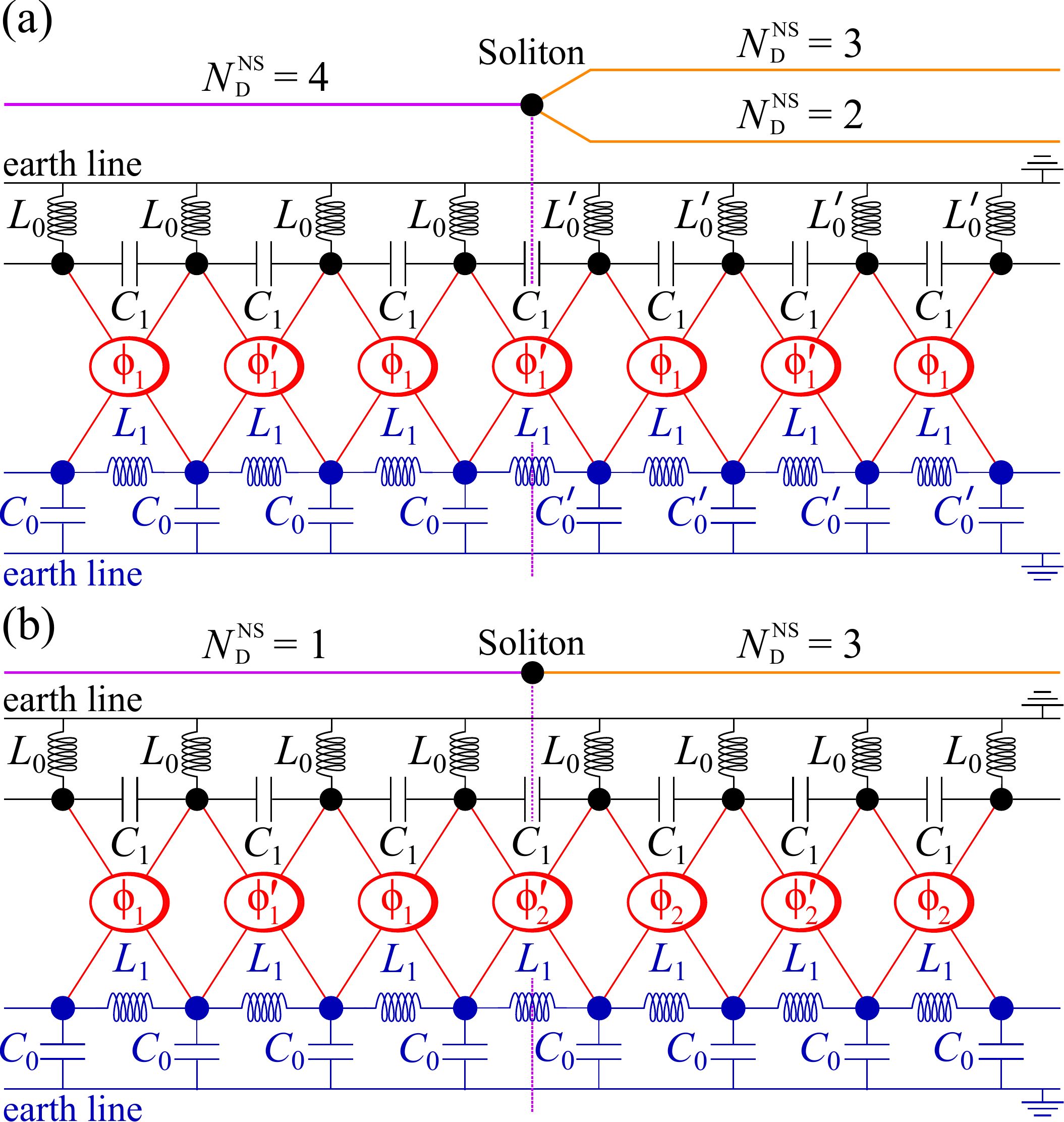}
    \caption{Examples of topolectric circuit realizations of the $\mathbb{Z}_4$ model. The model is simulated by two channels, the upper channel (black) represents the electron chain and the lower channel (blue) represents the hole chain. The chains are grounded by inductors (capacitors) in the electron (hole) chain, nodes in the same channel are connected by inductors $L_1$ and capacitors $C_1$, and the channels are paired to one another (red) by phase control units \cite{ezawa19,ezawa20}, Fig.~\ref{PCUs}. In (a) a soliton in the grounding inductance causes a phase transition from $N_\mathrm{D}^\mathrm{NS}=4$ to $N_\mathrm{D}^\mathrm{NS}=3$ or $N_\mathrm{D}^\mathrm{NS}=3$ depending on the parameter values. In (b) a soliton in the phase control unit causes a phase transition from $N_\mathrm{D}^\mathrm{NS}=1$ to $N_\mathrm{D}^\mathrm{NS}=3$. There may be different phase transitions to those shown here or no phase transition depending on the parameter values of the system.}
    \label{topoz4schematic}
\end{figure}
%%%%%%%%%%%%%%%%%%%%%%%%%%%%%%%%%%%%%%%%%%%%%%%%%%%%%%%%%

%%%%%%%%%%%%%%%%%%%%%%%%%%%%%%%%%%%%%%%%%%%%%%%%%%%%%%%%%
\begin{figure*}
    \centering
    \includegraphics[width=\linewidth]{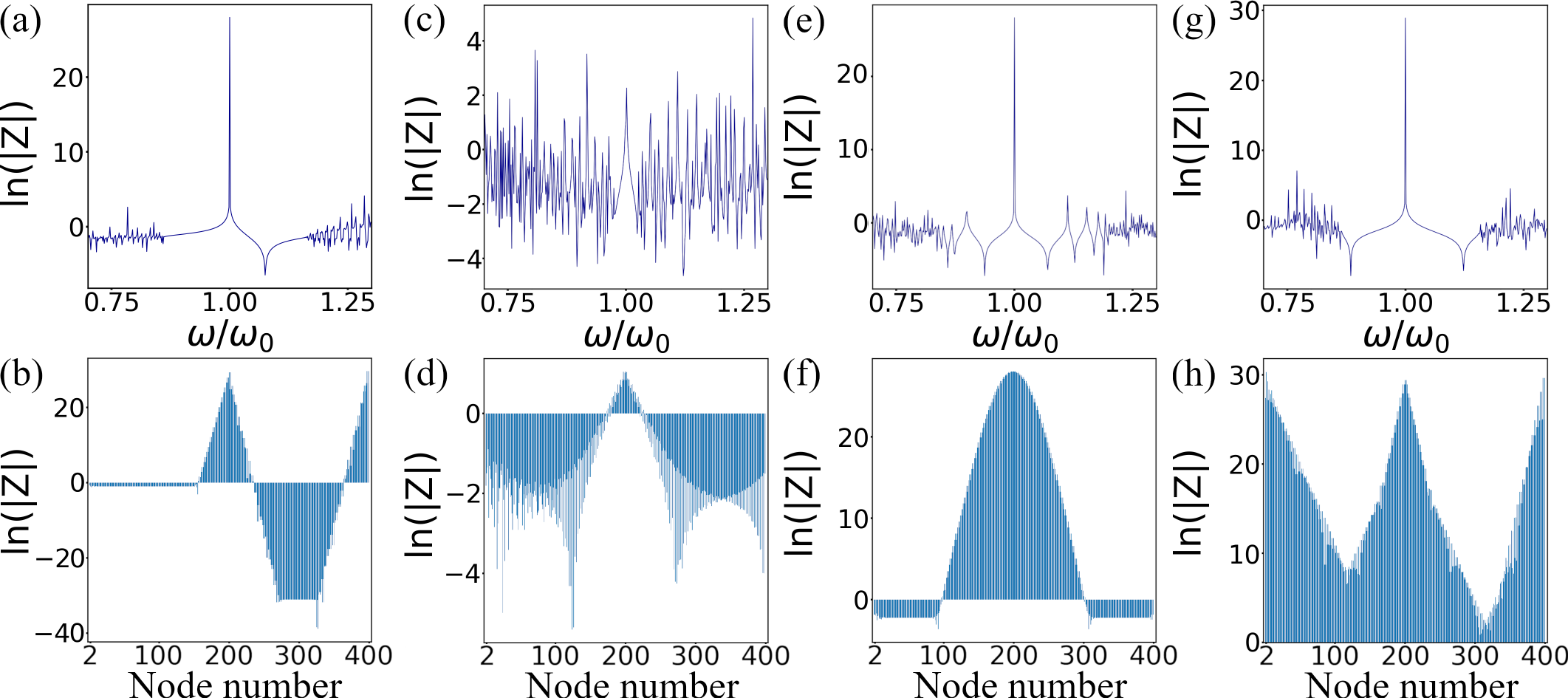}
    \caption{Impedances across the topolectric realization of the $\mathbb{Z}_4$ chain, Fig.~\ref{topoz4schematic}, for 400 nodes and input current at the first node in the electron channel. The top row of figures measure the impedance as a function of frequency with current output at the center of the chain, while the bottom row measures impedance as a function of variable output node. Figures (a)-(d) are for a system with an atomically sharp soliton in the grounding inductors and capacitors, where (a) and (b) are for a system with a soliton between phases $N_\mathrm{D}^\mathrm{NS}=4$ and $N_\mathrm{D}^\mathrm{NS}=3$, while (c) and (d) are for a soliton between phases $N_\mathrm{D}^\mathrm{NS}=4$ and $N_\mathrm{D}^\mathrm{NS}=2$. Figures (e) and (f) are for an atomically smooth soliton of width $\zeta=25$ in the grounding inductors and capacitors between phases $N_\mathrm{D}^\mathrm{NS}=4$ and $N_\mathrm{D}^\mathrm{NS}=2$. Finally, (g) and (h) are for an atomically sharp soliton in the phase-control unit between phases $N_\mathrm{D}^\mathrm{NS}=3$ and $N_\mathrm{D}^\mathrm{NS}=1$. Parameter values for (a) and (b) are $L_0=0.075$H, $C_0=1.6$F, $L_0^\prime=0.171$, $C_0^\prime=0.7$ $L_1=0.4$H, $C_1=0.3$F, $L_S=0.6$H, $C_S=0.2$F, $R_S=5\Omega$, for (c) and (d) $L_0=0.047$H, $C_0=1.7$F, $L_0^\prime=0.053$H, $C_0^\prime=1.5$F, $L_1=0.1$H, $C_1=0.8$F, $L_S=0.3$H, $C_S=0.267$F, $R_S=0.2\Omega$, for (e) and (f) $L_1=0.1$H, $C_1=0.8$F, $L_S=0.3$H, $C_S=0.267$F, $R_S=0.2\Omega$ and $\mu_\mathrm{max}=1$, and for (g) and (h)  $L_0=0.4$H, $C_0=0.2$F, $L_1=0.1$H, $C_1=0.8$F, $L_S=0.3$H, $C_S=0.267$F, $R_S=1\Omega$.}
    \label{z4impedances}
\end{figure*}
%%%%%%%%%%%%%%%%%%%%%%%%%%%%%%%%%%%%%%%%%%%%%%%%%%%%%%%%%

The topolectric realization of the $\mathbb{Z}_4$ model is also of a similar form to the circuit description of the Kitaev chain found in Ref.~\cite{ezawa20}, and by extension the topolectric AII model. The $\mathbb{Z}_4$ chain can be modeled with two coupled channels as shown in Fig.~\ref{topoz4schematic}, where the upper channel represents the `electron' nodes and the lower channel represents the `hole' nodes \cite{ezawa19,ezawa20}. The electron (hole) channel is grounded by inductors $L_0$ or $L_0^\prime$ (capacitors $C_0$ or $C_0^\prime$) depending on the phase of the model. Nodes within the electron (hole) channel are connected by capacitors $C_1$ (inductors $L_1$). In contrast to the Kitaev chain, the alternating phases of the $\mathbb{Z}_4$ model cannot be gauged away \cite{tymczyszyn24}, and must be accounted for when constructing a topolectric circuit. As a result, the inter-channel couplings and phases are determined by PCUs, Fig.~\ref{PCUs}, where now we also make use of $\phi_1^\prime$ and $\phi_2^\prime$ which flip the sign of the resistance, creating the required alternating phase. Phase transitions are controlled by a change in the grounding inductance from $L_0$ to $L_0^\prime$ and grounding capacitance from $C_0$ to $C_0^\prime$, Fig.~\ref{topoz4schematic}(a), or by a change in the PCU from $\phi_1$ ($\phi_1^\prime$) to $\phi_2$ ($\phi_2^\prime$), Fig.~\ref{topoz4schematic}(b). The resonant frequency is
\begin{equation}
    \omega_0\equiv 1/\sqrt{L_0C_0} = 1/\sqrt{L_1C_1}=1/\sqrt{L_SC_S}\,.
    \label{z4frequency}
\end{equation} 
\noindent At this frequency, we can compare the components to the parameters of the $\mathbb{Z}_4$ Hamiltonian, Eq.~(\ref{nsham}), as
\begin{eqnarray}
    \mu &=& -2C_1+C_0\,, \label{z4equivalencesmu}\\
    v &=& -C_1 \,,\\
    \Delta_s &=& \sqrt{C_S^2+L_SC_S/R_S^2}\,, \\[1.5pt]
    \phi_s &=& \arg\left[\eta_tC_S+i\sqrt{L_SC_S}/R_S\right]\,, \label{z4equivalencesmu2}  
\end{eqnarray}
where $C_0$ is interchangeable with $C_0^\prime$ and
\begin{equation}
    \eta_t=\begin{cases}
        1 & \mathrm{if} \hspace{1cm} \mathrm{PCU}=\phi_1\,, \\
        -1 & \mathrm{if} \hspace{1cm} \mathrm{PCU}=\phi_2\,,
    \end{cases}
    \label{etaval}
\end{equation}
assuming that the PCUs do not alternate along the chain as they do in the AII topolectrics. Substituting these expressions into the phase transitions shown in Fig.~\ref{z4fig1}(d) we find that
\begin{equation}
    N_\mathrm{D}^\mathrm{NS}=
    \begin{cases}
        1 & \mathrm{if} \hspace{2mm} X<0 \hspace{2mm}\mathrm{and} \hspace{2mm}Y>0\,,\\
        2 & \mathrm{if}\hspace{2mm} X>0 \hspace{2mm}\mathrm{and}\hspace{2mm} Z<0 \,,\\
        3 & \mathrm{if}\hspace{2mm} X<0\hspace{2mm}\mathrm{and}\hspace{2mm} Y<0\,,\\
        4 & \mathrm{if}\hspace{2mm} X>0\hspace{2mm}\mathrm{and}\hspace{2mm} Z>0 \,,
    \end{cases}
    \label{z4topphases}
\end{equation}
where
\begin{eqnarray}
    X&=&C_0^2-4C_1C_0+\frac{4L_SC_S}{R_S^2},\\[3.5pt]
    Y &=&\eta_t\left(\sqrt{\frac{L_S}{R^2_S}+C_S}\right)^{-1}\,\,\,,\\[3pt]
    Z &=&\mu= -2C_1+C_0\,.
\end{eqnarray}
We can select parameter values such that by swapping $C_0$ for $C_0^\prime$ we can flip the sign of $X$ or $Z$, resulting in a phase transition between $N_\mathrm{D}^\mathrm{NS}=4$ and $N_\mathrm{D}^\mathrm{NS}=3$ or between $N_\mathrm{D}^\mathrm{NS}=4$ and $N_\mathrm{D}^\mathrm{NS}=2$ at the position in the chain that this swap occurs, Fig.~\ref{topoz4schematic}(a). Similarly, we can swap the PCU $\phi_1$ ($\phi_1^\prime$) for $\phi_2$ ($\phi_2^\prime$) which flips the sign of $Y$, resulting in a transition between phases $N_\mathrm{D}^\mathrm{NS}=1$ and $N_\mathrm{D}^\mathrm{NS}=3$. We note that the solitons shown in Fig.~\ref{topoz4schematic} are example transitions, and, alternatively, there may be different phase transitions or no phase transition depending on the parameter values.

Taking finite chains, with topological transitions occurring at the center of the chain, results in zero-energy states hosted on solitons and corresponding peaks in the impedance that are localized onto the solitons according to Eq.~(\ref{2PI}). We first plot the impedance for a system with 400 nodes, current input at the first site in the `electron' channel, and a soliton between $N_\mathrm{D}^\mathrm{NS}=4$ and $N_\mathrm{D}^\mathrm{NS}=3$ as a function of frequency in Fig.~\ref{z4impedances}(a). Here a large peak is observed at the resonant frequency. For the same system we then plot the impedance as a function of the output node, Fig.~\ref{z4impedances}(b), where peaks are exponentially (linearly in the log scale) localized at the edges of the $N_\mathrm{D}^\mathrm{NS}=3$ section of the chain, which hosts MZM. By deforming the parameter values, the transition may then be between $N_\mathrm{D}^\mathrm{NS}=4$ and $N_\mathrm{D}^\mathrm{NS}=2$. As neither side of this transition is in a phase that hosts MZM, the only contribution to the impedance peak is from the soliton. Fig.~\ref{z4impedances}(c) shows that the on-resonance peak for current output on the soliton is of the same order of magnitude as the off-resonance impedances, although Fig.~\ref{z4impedances}(d) does show that there is some small localization on the soliton despite the small magnitude. To obtain a more measurable impedance, we instead construct an atomically smooth soliton with a texture in the grounding components $L_0$ and $C_0$ between phases $N_\mathrm{D}^\mathrm{NS}=4$ and $N_\mathrm{D}^\mathrm{NS}=2$. The texture is described by
\begin{eqnarray*}
    L_{0,l}&=&\left[\omega^2\left(2C_1-\mu_\mathrm{max}\tanh\left(\frac{l-N-1/2}{\zeta}\right)\right)\right]^{-1}\,,\\
    C_{0,l}&=&\frac{1}{\omega^2L_{0,l}}=2C_1-\mu_\mathrm{max}\tanh\left(\frac{l-N-1/2}{\zeta}\right)\,,
\end{eqnarray*}
where $l$ is the index of a given node in the electron chain and $\mu_\mathrm{max}$ is the value of the onsite energy, Eq.~(\ref{z4equivalencesmu}), at infinity. We plot the impedance in the presence of a smooth soliton with width $\zeta=N/4$ as a function of frequency in Fig.~\ref{z4impedances}(e). This shows a much more distinct peak when compared to the atomically sharp soliton in Fig.~\ref{z4impedances}(c), and the impedance as a function of output node shows the corresponding peak at the center of the chain. Finally, we plot the impedance for a system with a soliton in the superconducting phase, Fig.~\ref{topoz4schematic}(b), between phases $N_\mathrm{D}^\mathrm{NS}=3$ and $N_\mathrm{D}^\mathrm{NS}=1$ as a function of frequency in Fig.~\ref{z4impedances}(g) and as a function of output node in Fig.~\ref{z4impedances}(h). Large peaks in the impedance are found localized at both ends of the chain in addition to localization on the soliton at the center of the chain. This is due to both $N_\mathrm{D}^\mathrm{NS}=3$ and $N_\mathrm{D}^\mathrm{NS}=1$ hosting MZM.

\section{Disorder in the $\mathbb{Z}_{4}$ model}
\label{disordersec}

\subsection{Minimal model}

In general, nonsymmorphic symmetries are sensitive to disorder because they rely on translation symmetry, which is violated by sharp disorder. This is particularly relevant in topolectric realizations, where unavoidable component tolerances introduce variations along the chain, effectively acting as uncorrelated on-site or bond disorder \cite{lee18,yang20,wu20}. In practice, such nonuniformity both detunes and broadens the impedance signature associated with boundary and defect resonances. Similar broadening and spectral shifts can also arise from resistive losses, which render the effective circuit Hamiltonian weakly non-Hermitian \cite{lee18,liu20,helbig20,tang25}, although we do not analyze these effects explicitly here.

In this work, our goal is instead to isolate how disorder in specific model parameters impacts the unconventional topology of the $\mathbb{Z}_4$ model. Because the mapping of circuit components onto the effective pairing terms is nontrivial, we implement disorder at the level of the condensed-matter parameters and relate it back to the topolectric realization via Eqs.~(\ref{z4equivalencesmu})–(\ref{z4equivalencesmu2}).

%%%%%%%%%%%%%%%%%%%%%%%%%%%%%%%%%%%%%%%%%%%%%%%%%%%%%%%%%%%%%%%%
\begin{figure*}
    \centering
    \includegraphics[width=\linewidth]{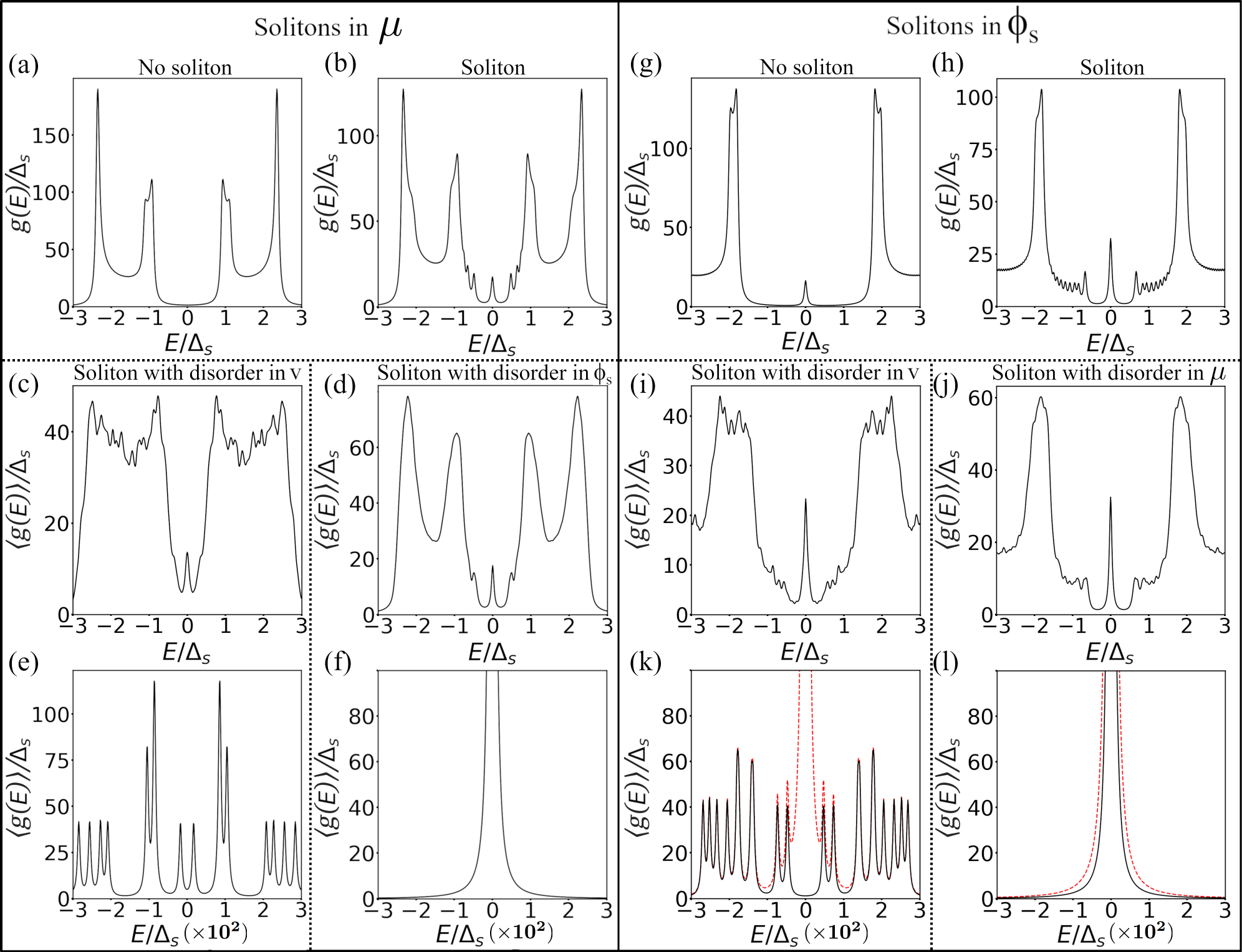}
    \caption{Density of states (DOS) for various systems of the $\mathbb{Z}_{4}$ model in position space with 48 unit cells. (a) The DOS in the phase $N_\mathrm{D}^\mathrm{NS}=4$ with $\mu=1$, $v=0.3$, $\Delta_s=1$ and $\phi_{s}=\pi/4$, (b) is for the same system but in the presence of a soliton in the chemical potential with $\mu_{\mathrm{max}}=1$ and width $\zeta=12$. Figures (c)-(f) detail the DOS once disorder has been introduced to (b). (c) and (d) show the DOS for atomically sharp disorder in $v$ and $\phi_{s}$, respectively. (e) and (f) show the same disorder as (c) and (d) but for small energy of order $\Delta_{s}\times10^{-2}$ for disorder in $v$ and $\phi_{s}$, respectively. (g) The DOS in the phase $N_\mathrm{D}^\mathrm{NS}=3$ with $\mu=1$, $v=1.5$, $\Delta_{s}=1$, and $\phi_{s}=\pi/4$, (h) is for same system but in the presence of a soliton in the superconducting phase with $\phi_{s,\mathrm{max}}=\pi$ and width $\zeta=12$. Figures (i)-(l) detail the DOS once disorder has been introduced to (h), (i) and (j) show atomically sharp disorder in $v$ and $\mu$, respectively. (k) and (l) show the same disorder as (i) and (j) for a small energy of order $\Delta_{s}\times10^{-2}$ in $v$ and $\mu$, respectively. A broadening width of $\xi=0.04\Delta_{s}$ was used for figures~(a)-(d) and (g)-(j), and $\xi=0.0004\Delta_{s}$ for figures (e)-(f) and (k)-(l). For plots with disorder we average over 20 disorder realizations with a disorder magnitude of $W=0.5$ and parameter $\eta=0.05$, resulting in strong atomically sharp disorder within each sample.}
    \label{dosmain}
\end{figure*}
%%%%%%%%%%%%%%%%%%%%%%%%%%%%%%%%%%%%%%%%%%%%%%%%%%%%%%%%%%%%%%%%

We introduce disorder in the chemical potential through the addition $\delta\mu_{l}$ to the $l$th atomic site in the particle chain, as described by the Gaussian-correlated potential \cite{koschny02,guo08,muten24} given by
\begin{equation}
    \delta\mu_{l}=\frac{\sum_{m}w_{m}\exp(-|l-m|^{2}/\eta^{2})}{\sqrt{\sum_{m}\exp(-|l-m|^{2}/\eta^{2})}},
    \label{disordereq}
\end{equation}
where $\eta$ is the correlation length in dimensionless units. The summation is over all atomic sites $m=1,2,...,2N$ with $w_{m}$ drawn randomly from a uniform distribution $-W\leq w_{m}\leq W$ with disorder strength $W$. This allows for atomically sharp disorder within a sample for the case of $\eta\ll 1$ and sample-to-sample variations for $\eta\gg N$ across an ensemble. We find that sample-to-sample disorder does not affect topologically protected zero modes in this model, as for each variation in the ensemble the symmetries are not broken, hence we focus on atomically sharp disorder within each member of the ensemble. In this way, chosen parameters can be effectively randomized across the length of the chain. This breaks the nonsymmorphic symmetry that relies on translational invariance, while maintaining the symmorphic charge-conjugation symmetry that protects the MZM localized onto the ends of the chain in phases $N_\mathrm{D}^{\mathrm{NS}}=3$ and $N_\mathrm{D}^{\mathrm{NS}}=1$. We can also use the right hand side of Eq.~(\ref{disordereq}) to calculate disorder terms for the hopping parameter $\delta v_{l}$ and the superconducting phase $\delta\phi_{s,l}$. We plot the density of states (DOS) for these finite systems numerically by approximation in Fig.~\ref{dosmain} using a Lorentzian of finite width $\xi$,
\begin{equation}
    g(E)=\frac{1}{\pi}\sum_{n}\frac{\xi}{(E-E_{n})^{2}+\xi^{2}}.
\end{equation}
We first plot the DOS with width $\xi=0.04\Delta_{s}$ for a system of 48 unit cells, with no solitons or MZM and no disorder. This system resides in the phase $N_\mathrm{D}^\mathrm{NS}=4$ with parameter values $\mu=1$, $v=0.3$, $\Delta_{s}=1$, and $\phi_{s}=\pi/4$ and is shown in Fig.~\ref{dosmain}(a). As expected the DOS is zero at zero energy, indicating the absence of any zero-energy states. Fig.~\ref{dosmain}(b) shows the DOS for the same system after a soliton in the chemical potential is added with $\mu_{\mathrm{max}}=1$. There is now a small but distinct peak in the DOS at zero energy, corresponding to a zero energy soliton exponentially localized at the center of the chain, Fig.~\ref{solwf}(a). While maintaining the soliton we now also introduce atomically sharp disorder with $\eta=0.05$: The disorder-averaged DOS are shown in Fig.~\ref{dosmain}(c) and Fig.~\ref{dosmain}(d) for disorder in $v$ and $\phi_{s}$, respectively. For all DOS plots with disorder we average over 20 disorder realizations. As the disorder breaks the translational invariance of the system and hence the nonsymmorphic symmetries, we expect that the soliton states would delocalize into the bulk and move away from zero energy. However, Fig.~\ref{dosmain}(c) and Fig.~\ref{dosmain}(d) clearly show a zero-energy peak in the DOS. 

To better understand the robustness of such soliton states we plot the disorder-averaged DOS for small energy $\sim\Delta_{s}\times 10^{-2}$ with disorder in $v$ and $\phi_{s}$. Fig.~\ref{dosmain}(e) shows that disorder in $v$ does delocalize the soliton state, increasing its energy away from zero. In contrast, Fig.~\ref{dosmain}(f) shows that disorder in $\phi_{s}$ does not delocalize the soliton, as the energy eigenvalues across the ensemble remain fixed at zero energy. 

We repeat this procedure for the case of solitons in the superconducting phase $\phi_{s}$ by plotting the DOS with $\xi=0.04\Delta_{s}$. We start with a system in the $N_\mathrm{D}^\mathrm{NS}=3$ phase with no soliton and parameter values $\mu=1$, $v=1.5$, $\Delta_{s}=1$, and $\phi_{s}=\pi/4$. In this phase there are MZM localized onto the edges of the chain even in the absence of solitons. Fig.~\ref{dosmain}(g) shows the DOS for this pristine system, where the MZM are shown as a small but observable peak at zero energy. Adding a soliton in $\phi_{s}$ between the phases $N_\mathrm{D}^\mathrm{NS}=3$ and $N_\mathrm{D}^\mathrm{NS}=1$, increases the magnitude of this peak, as can be seen in Fig.~\ref{dosmain}(h). Adding disorder to the hopping parameter $v$ and chemical potential $\mu$ does not appear to significantly change the DOS, as shown in Fig.~\ref{dosmain}(i) and Fig.~\ref{dosmain}(j), respectively. Fig.~\ref{dosmain}(k) and Fig.~\ref{dosmain}(l) show the disorder-averaged DOS for small energy $\sim\Delta_{s}\times 10^{-2}$ for disorder in $v$ and $\mu$, respectively. As we are focused on the effect of disorder on the soliton states we separate the MZM from the DOS by representing their contribution with the red dashed line in Fig.~\ref{dosmain}(k) and \ref{dosmain}(l). In Fig.~\ref{dosmain}(k), it is clear that the disorder in $v$ has delocalized the soliton state and moved its energy away from zero, while disorder in $\mu$ has not, with the soliton remaining at zero energy.

These results indicate an apparent channel-selective robustness in the minimal nearest-neighbor model, i.e., although open boundaries and atomically sharp disorder generically break translation and hence the nonsymmorphic symmetries, the domain-wall bound state can remain near zero energy for certain disorder channels without being protected by the nonsymmorphic bulk invariant. In contrast, the end Majorana zero modes (when present) are protected by the symmorphic particle–hole symmetry, whereas the near-zero domain-wall pinning here is an accidental property of the minimal Hamiltonian, where some symmetry-breaking perturbations couple only weakly within the soliton subspace. To see this, we can examine a first-order perturbation of the soliton energies. Take, for example, the soliton in the chemical potential $\mu$ with disorder in $\phi_s$, a robust disorder channel, such that we may describe the Hamiltonian as a function of the many values of disorder, $H(\mathbf{\delta\phi_{s}})$, where $\mathbf{\delta\phi_s}=(\delta\phi_{s,1},\delta\phi_{s,2},\ldots,\delta\phi_{s,2N-1})$. Performing an expansion around zero disorder, $\mathbf{\delta\phi_{s}}=0$, we find that
\begin{equation}
    H(\mathbf{\delta\phi_{s}})=H_0+\delta H(\mathbf{\delta\phi_{s}})+\mathcal{O}((\mathbf{\delta\phi_{s}})^2)+\ldots\,,
\end{equation}
where
\begin{equation}
    \delta H(\mathbf{\delta\phi_{s}}) = \sum_l\delta\phi_{s,l}\left.\frac{\partial H}{\partial(\delta\phi_{s,l})}\right|_{\delta\phi_{s,l}=0}\,.
\end{equation}
Therefore, the first order energy shifts of the soliton state are determined by the eigenvalues of the $2\times 2$ effective Hamiltonian $H_{\mathrm{eff}}=P\delta HP$, where $P$ is a projector onto the soliton subspace composed of the soliton related eigenvectors. This applies when the soliton subspace is spectrally isolated from the rest of the energy levels, so that degenerate perturbation theory is valid. We find that, for each term of $\delta H$, which each correspond to an individual noise value, the contribution to the effective Hamiltonian,
\begin{equation}
    M_l=P\left(\left.\frac{\partial H}{\partial(\delta\phi_{s,l})}\right|_{\delta\phi_{s,l}=0}\right)P\approx0\,,
\end{equation}
within numerical precision. Therefore, the soliton energy has no first-order shift due to none of the disorder terms coupling to the soliton subspace. A similar analysis for disorder in the hopping $v$, a non-robust disorder channel, shows that $M_l\neq0$ in general, resulting in a first-order energy shift of the soliton states. An analogous relation holds for solitons in $\phi_s$, although the effective Hamiltonian is a $4\times 4$ matrix due to the presence of two MZM in addition to the two soliton energy levels. While this is not definitive proof of the mechanism of robustness for certain disorder channels with large disorder strength, as higher-order terms may become non-negligible, it provides a qualitative explanation for the observed DOS calculations.

We find that adding symmetry-allowed longer-range terms removes this emergent constraint, allowing the bound-state energy to generically shift away from zero.

\subsection{Long-range parameters}
\label{HOPsec}

%%%%%%%%%%%%%%%%%%%%%%%%%%%%%%%%%%%%%%%%%%%%%%%%%%%%%%%%%%%%%%%%
\begin{figure}
    \centering
    \includegraphics[width=\linewidth]{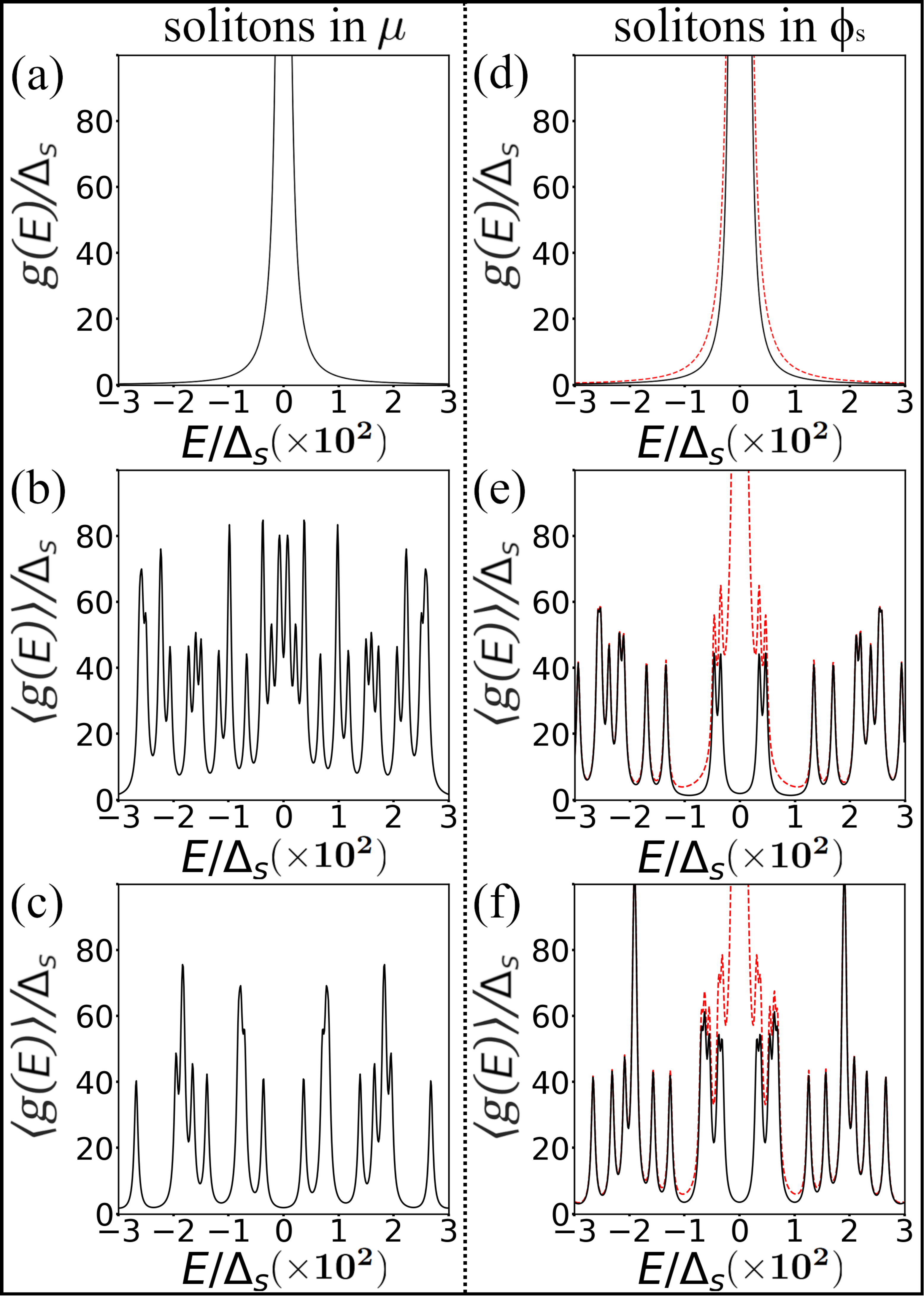}
    \caption{Density of states (DOS) for small energy of order $\Delta_{s}\times10^{-2}$ for various systems of the $\mathbb{Z}_{4}$ model in position space with 48 unit cells and in the presence of a $\Delta_{p}$ order parameter. All figures use a broadening width $\xi=4\Delta_s\times10^{-4}$. In each of these systems, solitons of width $\zeta=12$ are present, with figures (b), (c), (e), and (f) also having disorder of strength $W=0.5$ and parameter $\eta=0.05$, averaged over 20 disorder realizations. (a) DOS in the phase $N_\mathrm{D}^\mathrm{NS}=4$, with a soliton in the chemical potential and parameter values $\mu_{\mathrm{max}}=1$, $\mu_c=-0.255$, $v=0.3$, $\Delta_s=1$, $\phi_{s}=\pi/4$, and $\Delta_{p}=0.5$. (b) The same system as (a), but with disorder in the hopping parameter $v$. (c) DOS for the same system as (a), but in the presence of disorder in $\phi_{s}$. (d) The DOS in the phase $N_\mathrm{D}^\mathrm{NS}=3$, with a soliton in the superconducting phase and parameter values $\mu=1$, $v=1.5$, $\Delta_{s}=1$, $\phi_{s,\mathrm{max}}=\pi$, $\phi_{s,c}=2.02$, and $\Delta_{p}=0.5$. (e) The same system as (d), but with disorder in the hopping parameter $v$. (f) The same system as (d), but with disorder in $\mu$.}
    \label{DOS2}
\end{figure}
%%%%%%%%%%%%%%%%%%%%%%%%%%%%%%%%%%%%%%%%%%%%%%%%%%%%%%%%%%%%%%%%

So far we have only considered the minimal required parameters to realize the $\mathbb{Z}_4$ topology, and this has led to a surprising robustness of solitons localized on these transitions under certain disorder channels. There are two first-order parameters, not counting their complex phases, that we can consider adding that still satisfy the required symmetries. These are an A-A hopping $t_{AA}$ and a p-wave order parameter $\Delta_{p}$. Here, we focus on the addition of $\Delta_{p}$. If we consider the simple case of switching off the s-wave superconducting parameter $\Delta_{s}$ and switching on the p-wave parameter $\Delta_{p}$, we find that the $\mathbb{Z}_{4}$ model is reduced to a $\mathbb{Z}_{2}$ index due to the creation of unitary symmetries. However, unlike typical superconductors in the symmorphic D class such as the Kitaev chain \cite{kitaev01}, having both $\Delta_{s}$ and $\Delta_{p}$ be non-zero results in a distortion of the topological phase transitions between $N_\mathrm{D}^\mathrm{NS}=1$ and $N_\mathrm{D}^\mathrm{NS}=3$ and between $N_\mathrm{D}^\mathrm{NS}=2$ and $N_\mathrm{D}^\mathrm{NS}=4$, although, as expected, the Majorana phase transition remains unchanged. Introducing $\Delta_{p}$ to the Hamiltonian~(\ref{nsham}) keeps the noninteracting part the same but results in a new coupling matrix of
\begin{equation*}
    \hat{\Delta} (k) = \begin{pmatrix}
2i\Delta_{p} \sin(ka) & 2i\Delta_{s} \sin (ka/2 + \phi_{s}) \\
2i\Delta_{s} \sin (ka/2 - \phi_{s})  & 2i\Delta_{p} \sin(ka)
\end{pmatrix}.
\label{gf1}
\end{equation*}
Analytically deriving the topological transition points is nontrivial, and is discussed in detail in Appendix~\ref{hop}. In the same appendix we also include the derivation of topological transition points for a Hamiltonian that includes the hopping parameter $t_{AA}$. Here, we examine the effect of disorder on the DOS for systems with solitons, where the zero energy states are localized on domain walls corresponding to the new transition points once $\Delta_{p}$ coupling is accounted for. The disorder-averaged DOS for small energy $\sim\Delta_{s}\times10^{-2}$ is shown in Fig.~\ref{DOS2}. We first examine the case of solitons in $\mu$ between $N_\mathrm{D}^\mathrm{NS}=2$ and $N_\mathrm{D}^\mathrm{NS}=4$ as we previously did for a minimal model, Fig.~\ref{dosmain}(a)-(f), for a soliton localized on $\mu=0$. Due to the introduction of the parameter $\Delta_{p}$, the value of $\mu$ which corresponds to a phase transition has shifted. For a system with $\Delta_{p}=0.5$, $v=0.3$, $\Delta_{s}=1$, and $\phi=\pi/4$, we find that the phase transition occurs at $\mu\approx-0.255$. This can be accounted for in the soliton texture, Eq.~(\ref{onsitesol}), by setting $\mu_{c}=-0.255$. In this way the zero-energy state remains localized on the soliton at the center of the chain. The disorder-averaged DOS for this system without disorder is shown in Fig.~\ref{DOS2}(a), and for disorder in $v$ and $\phi_{s}$ in Fig.~\ref{DOS2}(b) and Fig.~\ref{DOS2}(c), respectively. Similarly to the minimal model with disorder in $v$, Fig.~\ref{dosmain}(e), the DOS for a system with non-zero $\Delta_{p}$ and disorder in $v$ has multiple peaks near zero energy, rather than a single larger peak at exactly zero energy. However, in the case of the minimal model we found the soliton state to be especially robust to the presence of disorder in $\phi_{s}$, Fig.~\ref{dosmain}(f), remaining at zero energy. In contrast to this, the presence of $\Delta_{p}$ shifts the states from zero energy, Fig.~\ref{DOS2}(b), as there are no longer any emergent constraints at the soliton.  

%%%%%%%%%%%%%%%%%%%%%%%%%%%%%%%
\begin{figure}
    \centering
    \includegraphics[width=\linewidth]{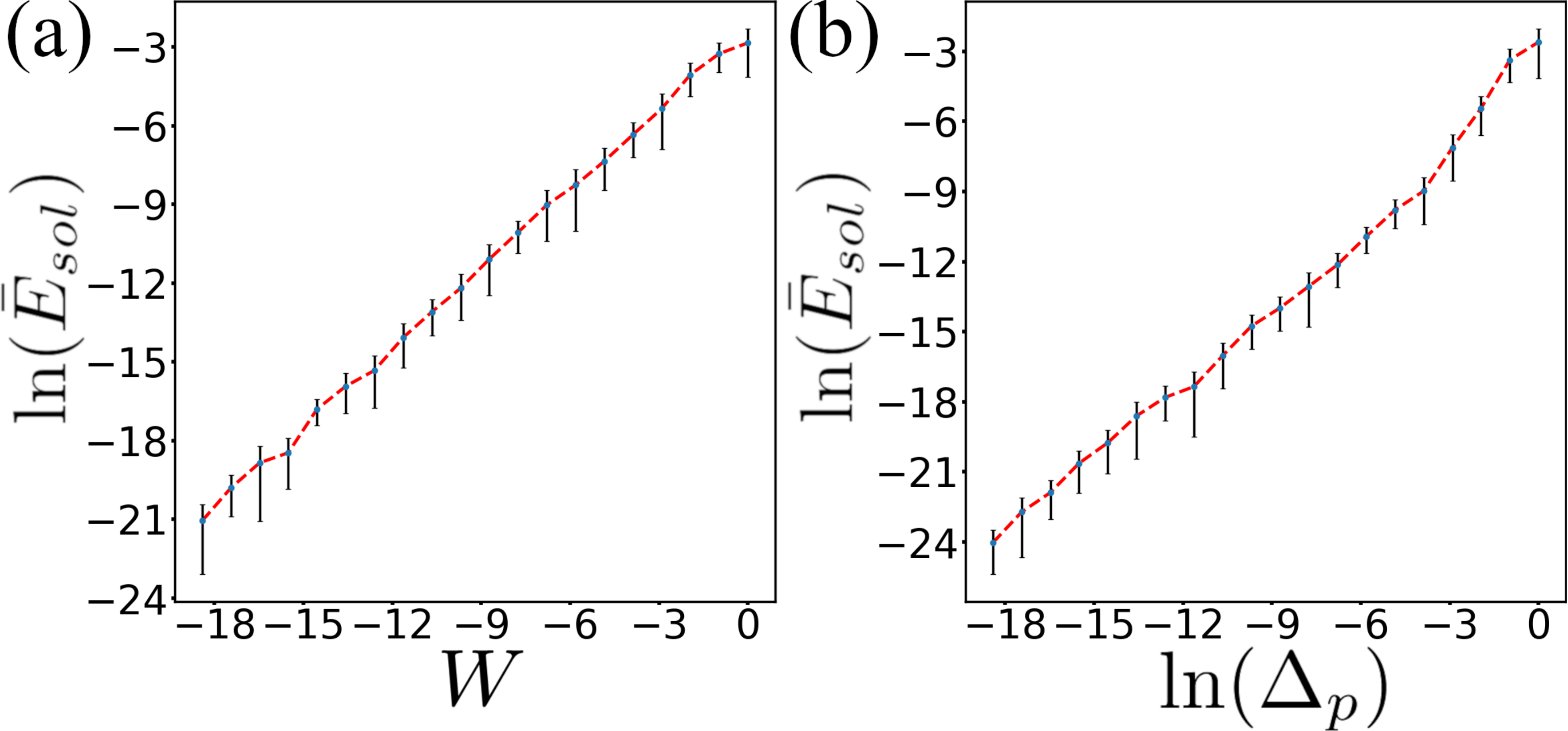}
    \caption{Disorder-averaged soliton energy for a soliton in the chemical potential $\mu$ of width $\zeta=12$ in the $\mathbb{Z}_{4}$ model between phases $N_\mathrm{D}^\mathrm{NS}=2$ and $N_\mathrm{D}^\mathrm{NS}=4$. Disorder is added to the superconducting phase with parameter $\eta=0.05$. (a) Energy plotted against the disorder strength $W$ with fixed parameter $\Delta_p=0.5$, where the energy is represented on a logarithmic scale. (b) Energy plotted against the magnitude of the superconducting order parameter $\Delta_{p}$, with fixed disorder strength $W=0.5$ and represented as a log-log plot. In both (a) and (b) each point is averaged across an ensemble of 20 disorder realizations of 48 unit cells. Error bars represent ranges of $\pm1$ standard deviation from the mean for each ensemble. The plot shows approximately linear behavior for both (a) and (b) indicating an exponential relation in (a) and a power-law relation in (b). Parameter values are $\mu_{\mathrm{max}}=1$, $\Delta_s=1$, $v=0.3$, $\phi_s=\pi/4$, and in (a) $\mu_c=-0.255$ while for (b) $\mu_c$ is variable depending on the value of $\Delta_p$.}
    \label{AVEvsDP}
\end{figure}
%%%%%%%%%%%%%%%%%%%%%%%%%%%%%%%

We plot the logarithm of the disorder-averaged soliton energy $\ln(\bar{E}_\mathrm{sol})$ for a soliton in the chemical potential $\mu$ against the disorder strength $W$ in Fig.~\ref{AVEvsDP}(a), and against the log value of $\Delta_{p}$ in Fig.~\ref{AVEvsDP}(b), with error bars showing $\pm1$ standard deviation from the mean. Each point is averaged across an ensemble of 20 disorder realizations for disorder in the superconducting phase $\phi_{s}$ with fixed $\Delta_p=0.5$ in Fig.~\ref{AVEvsDP}(a) and fixed disorder strength $W=0.5$ in Fig.~\ref{AVEvsDP}(b). The approximately linear behavior in both plots indicates an exponential relation between disorder strength and the raising of the soliton energy level from zero energy, and a power-law relation between the value of $\Delta_{p}$ and the raising of the soliton energy level from zero energy.

We also examine solitons in $\phi_{s}$ between phases $N_\mathrm{D}^\mathrm{NS}=1$ and $N_\mathrm{D}^\mathrm{NS}=3$, as we previously did for a minimal model as shown in Fig.~\ref{dosmain}(g)-(l) for a soliton localized on $\phi_{s}=\pi/2$. For a system with $\Delta_{p}=0.5$, $v=1.5$, $\Delta_{s}=1$, and $\mu=1$, we find that the phase transition occurs at $\phi_{s} \approx 2.02$. This can be accounted for in the soliton texture, Eq.~(\ref{SPsol}), by setting $\phi_{s,c}=2.02$ such that the zero energy state remains localized on the soliton at the center of the chain. The disorder-averaged DOS for this system for no disorder is shown in Fig.~\ref{DOS2}(d), and for disorder in $v$ and $\mu$ in Fig.~\ref{DOS2}(e) and Fig.~\ref{DOS2}(f), respectively. Similarly to the minimal model with disorder in $v$, Fig.~\ref{dosmain}(k), the DOS for a system with non-zero $\Delta_{p}$ and disorder in $v$ has multiple peaks near zero energy, rather than a single larger peak at exactly zero energy. However, in the case of the minimal model we found the soliton state to be especially robust to the presence of disorder in $\mu$, Fig.~\ref{dosmain}(l), remaining at zero energy. In contrast to this, the presence of $\Delta_{p}$ shifts the states from zero energy, Fig.~\ref{DOS2}(d).

\section{Conclusions}

We describe a method of calculating the topological index of nonsymmorphic one-dimensional tight-binding models with Kramers degeneracy by way of a generalized winding number. The method is demonstrated for superconducting models that realize two distinct symmetry implementations. In the nonsymmorphic AII class, time-reversal symmetry is symmorphic while charge-conjugation and chiral symmetries are nonsymmorphic, resulting in a $\mathbb{Z}_2$ topological index. In the nonsymmorphic D class, charge-conjugation symmetry is symmorphic while time-reversal and chiral symmetries are nonsymmorphic, and the corresponding topology is characterized by a $\mathbb{Z}_4$ invariant. In both settings, the resulting phase structure is consistent with the nonsymmorphic classification of Ref.~\cite{shiozaki16} for nonsymmorphic models. While the topology of other symmetry classes can be described as either topologically trivial, an integer $\mathbb{Z}$ winding number, or a $\mathbb{Z}_{2}$ index, the $\mathbb{Z}_{4}$ index is unique in one-dimensional noninteracting models. We calculate the topology based on a nonsymmorphic winding number equivalent that has previously only been used to calculate a $\mathbb{Z}_2$ topological index without Kramers degeneracy \cite{shiozaki15,brzezicki20}. We explore the $\mathbb{Z}_{4}$ model in detail, introducing solitons into the system and, by plotting the disorder-averaged DOS, we find that, in general, the zero-energy states hosted by the solitons are not robust to disorder. However, for only nearest-neighbor parameters, emergent constraints are present around the domain wall that stop the system from being gapped, these constraints are lifted in the presence of longer-range parameters.

A theoretical realization of nonsymmorphic topology is shown in the form of topolectric circuits, where electrical components are used to mimic the position space Hamiltonian, and the impedance through the circuit is used to identify domain walls between degenerate ground states \cite{ezawa18,ezawa19,ezawa20}. To introduce nonsymmorphic topolectrics, we built circuits for the symmorphic SSH chains and compared them to the nonsymmorphic CDW chain. We then constructed topolectric equivalents of the previously described AII and $\mathbb{Z}_4$ models, finding agreement with the topological $k$-space calculations. We note that extensive experimental studies have been conducted on topolectric circuits \cite{haydar25}, in addition to finding analytic solutions for edge-to-edge impedances using the method of images \cite{haydar23}; these approaches may be extendable to the nonsymmorphic systems described here.

The models presented here in the atomic basis in position space provide a guideline for experimental realization of each of the two symmetry classes with four energy bands. Although the topology of the superconducting $\mathbb{Z}_4$ model may be realized by identical four-band noninteracting tight-binding parameters, it may be more experimentally viable to consider a two-band topological insulator, with superconductivity induced by the proximity effect \cite{franceschi10,albrecht16,nikolaenko21} to replicate the topology.

All relevant data presented in this paper can be accessed \cite{data}.

%%%%%%%%%%%%%%%%%%%%%%%%%%%%%%%%%%%%%%%%%%%%%%%%%%

\appendix

\section{Position space representations of AII and $\mathbb{Z}_4$ models}
\label{posmodels}

In order to model domain walls between topological phases we construct position space Hamiltonians for the AII and $\mathbb{Z}_4$ models, described in $k$-space by Eq.~(\ref{HAII}) and Eq.~(\ref{nsham}), respectively. The nonunitary symmetry operations can be written in position space as
\begin{align}
    \mathrm{time:}\hspace{1cm} & \mathcal{T}^{\dagger}H^{\ast}\mathcal{T} = H; \,\,\,\,\, \mathcal{T}\mathcal{T}^{\ast}=\pm I ;\\
    \mathrm{charge:}\hspace{1cm} & \mathcal{C}^{\dagger}H^{\ast}\mathcal{C} = -H; \,\,\,\,\, \mathcal{C}\mathcal{C}^{\ast}=\pm I ;\\
    \mathrm{chiral:}\hspace{1cm} & \mathcal{S}^{\dagger}H\mathcal{S} = -H; \,\,\,\,\, \mathcal{S}\mathcal{S}= I ;
\end{align}
where $\mathcal{T}$, $\mathcal{C}$, and $\mathcal{S}$ are unitary matrices, and $\mathcal{S}=\mathcal{T}^{\ast}\mathcal{C}$. For eight orbitals in the `ladder' basis $\Psi^{\dagger} = 
    \begin{pmatrix}
        c_{A,1}^{\dagger} & c_{B,1}^{\dagger} & c_{C,1}^\dagger & c_{D,1}^\dagger & c_{A,2}^{\dagger} & c_{B,2}^{\dagger} & c_{C,2}^\dagger & c_{D,2}^\dagger
    \end{pmatrix}$ the AII Hamiltonian may be written as
\begin{equation}
    \mathcal{H}=\Psi^{\dagger}\begin{pmatrix}
        \hat{h}_1 & \hat{\Delta}_1 & \hat{h}_2 & \hat{\Delta}_2 \\
        \hat{\Delta}_1^\dagger & -\hat{h}^\ast_1 & -\hat{\Delta}^\ast_2 & -\hat{h}_2^\ast \\
        \hat{h}_2^\dagger & -\hat{\Delta}_2^T & \hat{h}_1 & \hat{\Delta}_1 \\
        \hat{\Delta}_2^\dagger & -\hat{h}_2^T & \hat{\Delta}_1^\dagger & -\hat{h}_1^\ast
    \end{pmatrix}\Psi\,,
    \label{posspaceH}
\end{equation}
where
\begin{align*}
    \hat{h}_1\!&=\!\begin{pmatrix}
        u & v\\
        v & -u
    \end{pmatrix},\!\!\!\!&\hat{\Delta}_1\!=\!\begin{pmatrix}
        0 & \Gamma e^{i\phi_{\Gamma}}\\
        -\Gamma e^{i\phi_{\Gamma}} & 0
    \end{pmatrix},\\
    \hat{h}_2\!&=\!\begin{pmatrix}
        0 & 0\\
        v & 0
    \end{pmatrix},\!\!\!&\hat{\Delta}_2\!=\!\begin{pmatrix}
        0 & 0\\
        -\Gamma e^{-i\phi_{\Gamma}} & 0
    \end{pmatrix}.
\end{align*}
Nonsymmorphic symmetries must include a translation of half a unit cell, which, in this basis, can be written as
\begin{equation}
    T_{a/2} = 
    \begin{pmatrix}
        0 & 1 & 0 & 0 & 0 & 0 & 0 & \cdots & 0 & 0 & 0 \\
        0 & 0 & 0 & 0 & 1 & 0 & 0 & \cdots & 0 & 0 & 0 \\
        0 & 0 & 0 & 1 & 0 & 0 & 0 & \cdots & 0 & 0 & 0 \\
        0 & 0 & 0 & 0 & 0 & 0 & 1 & \cdots & 0 & 0 & 0 \\
        0 & 0 & 0 & 0 & 0 & 1 & 0 & \cdots & 0 & 0 & 0 \\
        \vdots & \vdots & \vdots & \vdots & \vdots & \vdots & \vdots & \vdots & \vdots & \vdots & \vdots \\
        0 & 0 & 0 & 0 & 0 & 0 & 0 & \cdots & 0 & 1 & 0 \\
        0 & 0 & 0 & 0 & 0 & 0 & 0 & \cdots & 1 & 0 & 0 \\
        1 & 0 & 0 & 0 & 0 & 0 & 0 & \cdots & 0 & 0 & 0 \\
        0 & 0 & 0 & 0 & 0 & 0 & 0 & \cdots & 0 & 0 & 1 \\
        0 & 0 & 1 & 0 & 0 & 0 & 0 & \cdots & 0 & 0 & 0 \\
    \end{pmatrix} \,.
\end{equation}
For this model, we find that $U_T=S_{yI}$, $U_C = T_{a/2}S_{zz}$, and $U_S=S_{xI}T_{a/2}S_{Iz}$, such that $U_C$ and $U_S$ are nonsymmorphic symmetry operators, where
\begin{equation}
    S_{ij} = 
    \begin{pmatrix}
        \tau_{i}\sigma_{j} & 0 & \cdots \\
        0 & \tau_{i}\sigma_{j} &\cdots \\
        \vdots & \vdots & \ddots
    \end{pmatrix}.
\end{equation}
For the $\mathbb{Z}_4$ model, the position space Hamiltonian can be written for a finite system with $J$ orbitals as a $2J\times 2J$ BdG matrix, e.g., for four orbitals in the `ladder' basis $\Psi^{\dagger} = 
    \begin{pmatrix}
        c_{A,1}^{\dagger} & c_{B,1}^{\dagger} & c_{A,1} & c_{A,1} & c_{A,2}^{\dagger} & c_{B,2}^{\dagger} & c_{A,2} & c_{A,2}
    \end{pmatrix}$ the Hamiltonian is
\begin{equation}
    \mathcal{H}=\frac{1}{2}\Psi^{\dagger}\begin{pmatrix}
        \hat{h}_1 & \hat{\Delta}_1 & \hat{h}_2 & \hat{\Delta}_2 \\
        \hat{\Delta}_1^\dagger & -\hat{h}^\ast_1 & -\hat{\Delta}^\ast_2 & -\hat{h}_2^\ast \\
        \hat{h}_2^\dagger & -\hat{\Delta}_2^T & \hat{h}_1 & \hat{\Delta}_1 \\
        \hat{\Delta}_2^\dagger & -\hat{h}_2^T & \hat{\Delta}_1^\dagger & -\hat{h}_1^\ast
    \end{pmatrix}\Psi+\mathrm{Tr}\,[\,\hat{h}_1\,],
\end{equation}
where
\begin{align*}
    \hat{h}_1\!&=\!\begin{pmatrix}
        -\mu & v\\
        v & -\mu
    \end{pmatrix},\!\!\!\!&\hat{\Delta}_1\!=\!\begin{pmatrix}
        0 & \Delta_{s}e^{i\phi_{s}}\\
        -\Delta_{s}e^{i\phi_{s}} & 0
    \end{pmatrix},\\
    \hat{h}_2\!&=\!\begin{pmatrix}
        t_{AA} & 0\\
        v & t_{AA}
    \end{pmatrix},\!\!\!&\hat{\Delta}_2\!=\!\begin{pmatrix}
        \Delta_{p} & 0\\
        \Delta_{s}e^{-i\phi_{s}} & \Delta_{p}
    \end{pmatrix},
\end{align*} 
where we have also included the longer-range order parameter $\Delta_p$ and hopping $t_{AA}$. The corresponding symmetry operators are $U_T=T_{a/2}$, $U_C=S_{xI}$, and $U_S=S_{xI}T_{a/2}$.

\section{Calculation of $\mathbb{Z}_{4}$ index in the presence of long-range parameters}
\label{hop}

The Hamiltonian of the $\mathbb{Z}_4$ model, Eq.~(\ref{nsham}), may be modified by the inclusion of long-range parameters $\Delta_p$ and $t_{AA}$, which represent a $p-$wave order parameter and next-nearest-neighbor hopping, respectively. The resulting Hamiltonian is
\begin{eqnarray}
{\cal H} (k) \!&=&\!\! \begin{pmatrix}
\hat{h} (k) & \hat{\Delta} (k) \\
\hat{\Delta}^{\dagger} (k) & -\hat{h}^{T} (-k)
\end{pmatrix} ,\label{hoham} \\
\hat{h} (k) \!&=&\!\! \begin{pmatrix}
- \mu+2t_{AA}\cos(ka) & 2 v \cos (ka/2) \\
2 v \cos (ka/2)  & -\mu+2t_{AA}\cos(ka)
\end{pmatrix} ,\nonumber \\
\hat{\Delta} (k) \!&=&\!\! \begin{pmatrix}
2i\Delta_{p} \sin(ka) & \!\!\!2i\Delta_{s} \!\sin (\!ka/2 \!+\! \phi_{s}\!) \!\\
\!2i\Delta_{s} \!\sin (\!ka/2 \!-\! \phi_{s}\!)  & 2i\Delta_{p} \sin(ka)
\end{pmatrix}\hspace{-0.5mm}.\hspace{5pt}\nonumber
\end{eqnarray}
We find the resulting Q-matrix (Eq.~(23) in the main text) to be $Q(k)=U_{xx}^{\dagger}\mathcal{H}(k)U_{xx}$ with
\begin{equation}
    q(k)=
    \begin{pmatrix}
        f(k) & g(k)\\
        g(k) & f^\ast(-k)
    \end{pmatrix},
    \label{Sqkappend}
\end{equation}
where
\begin{eqnarray}
    f(k)&=&\mu-2i\Delta_{s}\sin(ka/2+\phi_{s})-2t_{AA}\cos(ka),\\
    g(k) &=&-2v\cos(ka/2)+2i\Delta_{p}\sin(ka)\,.
\end{eqnarray}
From this form the required topology may be calculated by setting either $\Delta_p=0$ or $t_{AA}=0$. It is valid to consider both as non-zero, however, analytical complexity increases rapidly with the number of parameters.

\subsection{Superconducting order parameter $\Delta_{p}$}
\label{Sdpappendix}

The introduction of $\Delta_p$s significantly complicates the analytical calculation of the phase transition points between phases $N_\mathrm{D}^\mathrm{NS}=1$ and $N_\mathrm{D}^\mathrm{NS}=3$, and between phases $N_\mathrm{D}^\mathrm{NS}=2$ and $N_\mathrm{D}^\mathrm{NS}=4$, however, the Majorana boundary remains unchanged from the minimal model. To see this, we can consider the off-block diagonal Hamiltonian~(\ref{Sqkappend}) with $t_{AA}=0$. We again characterize the topology by plotting the product of the eigenvalues of $q(k)$ in the complex plane across the Brillouin zone, i.e. for $-\pi\leq ka<\pi$, where, for this system, the product can be written as
\begin{eqnarray}
    E_{q}(k)&=&\, 4\cos^2(ka/2)[4\Delta_{p}^2\sin^2(ka/2)\!+\!4i\Delta_{p}v\sin(ka/2)\nonumber\\
    &&+\Delta_{s}^{2}-v^2]+4i\Delta_{s}\mu\sin(ka/2)\cos(\phi_{s})\nonumber\\
    &&-4\Delta_{s}^{2}\cos^{2}(\phi_{s})+\mu^2.
    \label{SEqZ4nm}
\end{eqnarray}
From this, the Majorana number can be found by setting $k=0$ and determining whether $E_{q}(0)$ is greater than or less than zero. As expected this yields the same Majorana phase transition as the minimal model. The other phase transitions are not found as simply as those described for the minimal model, as the complex part of $E_{q}(k)$ is no longer zero across the Brillouin zone in the cases of $\mu=0$ or $\cos(\phi_{s})=0$, such that we must consider the value of $k$ when calculating the phase transition points. 

To calculate the topology we can utilize the symmetry of $E_{q}(k)$ about the real axis and consider only one half of the path, between $k=0$ and $k=\pi$. Separating Eq.~(\ref{SEqZ4nm}) into its complex and imaginary parts we can recognize that both these parts must be equal to zero at the phase transition as they pass through the origin, i.e. $\mathrm{Im}[E_q(k_{T})]=0$ and $\mathrm{Re}[E_q(k_T)]=0$, where $k_T$ is the value of $k$ at the transition point. As we are only interested in a transition that is dependent on the physical parameters of the system, we rearrange both $\mathrm{Im}[E_q(k_{T})]=0$ and $\mathrm{Re}[E_q(k_T)]=0$ for $k_{T}$ and equate them to one another. Rearranging for $k_{T}$ results in multiple solutions which arise from its periodicity, but discarding solutions with negative $k$ values and squaring each of these solutions results in two equations that can be written as $S_{\pm}=0$, where
\begin{equation}
    S_\pm\!=\!\frac{\!-2\Delta_s^2\!+\!8\Delta_{p}^2\!+\!2v^2\!\pm\!2\sqrt{f}}{2\Delta_s^2\!+\!8\Delta_{p}^2\!-\!2v^2\!\mp\!2\sqrt{f}}
    \!+\!\frac{\Delta_{s}\mu\cos(\phi_{s})\!+\!4\Delta_{p}v}{\Delta_{s}\mu\cos(\phi_{s})},
    \label{Sdpsol}
\end{equation}
where
\begin{eqnarray}
    f &=&\! -16\Delta_{s}^2\cos^2(\phi_s)\Delta_{p}^2\!+\!16\Delta_p^4\!+\!(8\Delta_{s}^2\!+\!4\mu^2\!\!-8v^2)\Delta_{p}^2\nonumber\\
    &&+(v-\Delta_{s})^2(v+\Delta_{s})^2.
\end{eqnarray}
Solutions to these equations that correspond to real $k$ values in the range $0\leq k<\pi$ correspond to points of topological phase transition between phases $N_\mathrm{D}^\mathrm{NS}=2$ and $N_\mathrm{D}^\mathrm{NS}=4$, and between phases $N_\mathrm{D}^\mathrm{NS}=1$ and $N_\mathrm{D}^\mathrm{NS}=3$ in the $\mathbb{Z}_4$ model. In the limit $\Delta_p\rightarrow0$, these solutions approach the minimal model transitions of $\mu=0$ and $\cos\phi_s=0$ depending on whether the Majorana number is 1 or -1. We find that
\begin{equation}
    S_\pm=\begin{cases}
        S_+ &\mathrm{if}\hspace{0.3cm}\lim_{k\rightarrow k_{T}}\mathrm{Re[}E_{q}(k)]=0^{+}\,,\\
        S_- &\mathrm{if}\hspace{0.3cm}\lim_{k\rightarrow k_{T}}\mathrm{Re}[E_{q}(k)]=0^{-}\,,
    \end{cases}
    \label{Sapproachcondition}
\end{equation}
while traversing the path from $k=0$ to $k=\pi$. We note that an identical solution can be obtained through the integral calculation described in Ref.~\cite{tymczyszyn24}, by tracking the path of the integral and calculating at which point it first crosses the discontinuity at the origin.

\subsection{Hopping parameter $t_{AA}$}
\label{Staaappendix}

%%%%%%%%%%%%%%%%%%%%%%%%%%%%%%%%%%%%%%%%%%%%%%%%%%%%%%%%%%%%%%%%
\begin{figure}
    \centering
    \includegraphics[width=\linewidth]{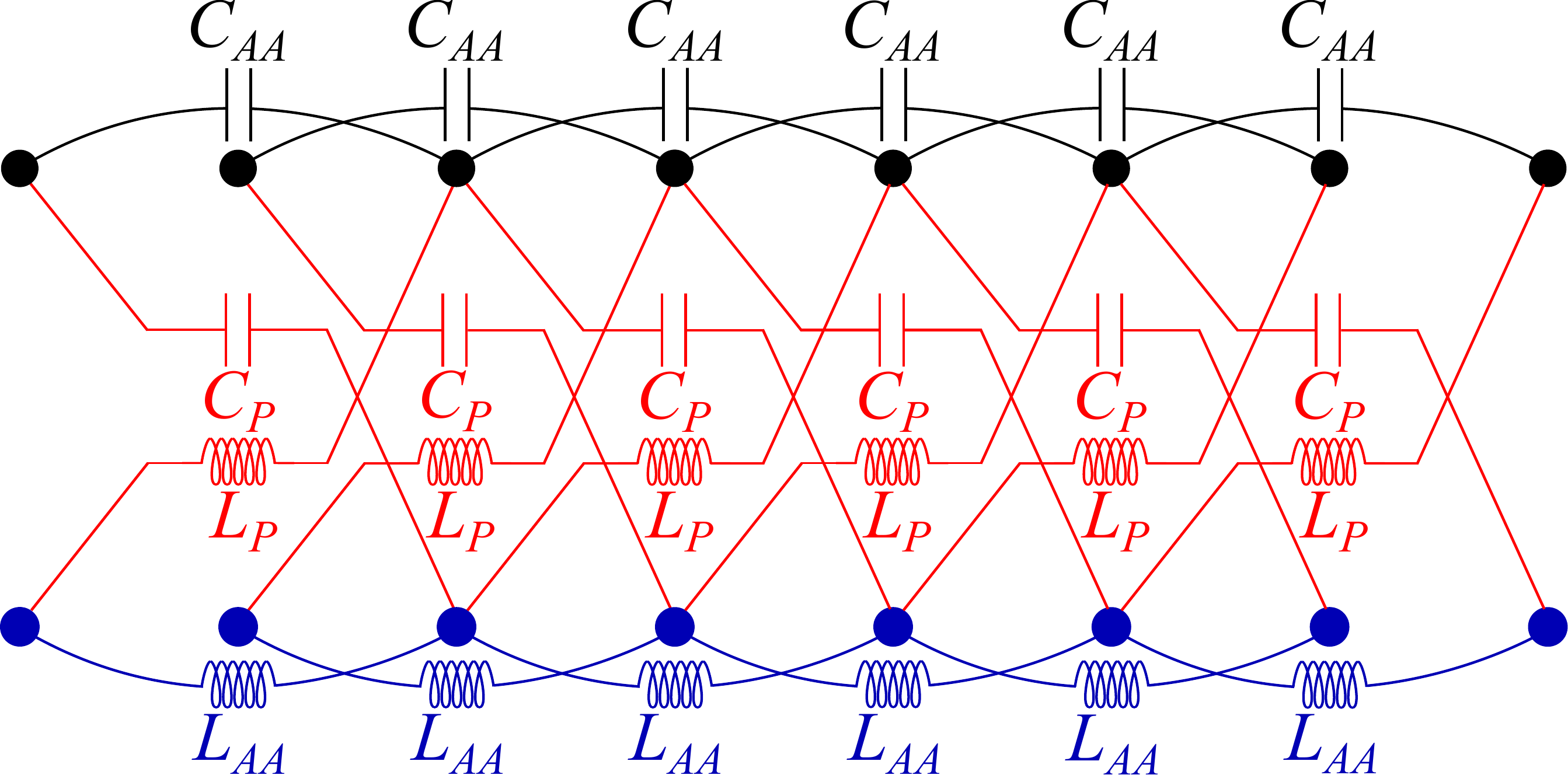}
    \caption{Topolectric circuit realization for the condensed matter parameters $\Delta_p$ and $t_{AA}$ in the $\mathbb{Z}_4$ model. The model is simulated by two channels, the upper channel (black) represents the electron chain and the lower channel (blue) represents the hole chain. The capacitors $C_{AA}$ in the upper channel represent $t_{AA}$, with the inductors $L_{AA}$ in the hole channel acting as their charge-conjugation partner. The two channels are paired (red) with inductors $L_P$ and capacitors $C_P$, which emulate the order parameter $\Delta_p$. Earth lines and grounding terms have been omitted, but can be found in Fig.~\ref{topoz4schematic}.}
    \label{z4longrangetopo}
\end{figure}
%%%%%%%%%%%%%%%%%%%%%%%%%%%%%%%%%%%%%%%%%%%%%%%%%%%%%%%%%%%%%%%%

Using a similar methodology as we did for $\Delta_{p}$ we now calculate the topological phase transitions in the presence of an additional hopping parameter between same sites in neighboring unit cells $t_{AA}$. To see this, we can consider the off-block diagonal Hamiltonian~(\ref{Sqkappend}) with $t_{AA}=0$. We again characterize the topology by plotting the product of the eigenvalues of $q(k)$ in the complex plane across the Brillouin zone, i.e. for $-\pi/a\leq k<\pi/a$, where, for this system, the product can be written as
\begin{eqnarray}
    E_{q}(k)&=&16t_{AA}\cos^4(ka/2)+4\cos^2(ka/2)[\Delta_{s}^2\!-\!v^2\nonumber\\
    &&-\!2\mu t_{AA}\!-\!4t_{AA}^2\!-\!4i\Delta_{s}t_{AA}\sin(ka/2)\cos(\phi_{s})] \nonumber\\
    &&-4\Delta_{s}\cos^2(\phi_{s})\!+\!(\mu+2t_{AA})^2\nonumber\\
    &&+4i\Delta_{s}\cos(\phi_{s})(\mu+2t_{AA})\sin(ka/2).
\end{eqnarray}
Unlike for the addition of parameter $\Delta_{p}$, we find that $t_{AA}$ does change the Majorana number transition. This transition can be found by setting $E_{q}(0)=0$, resulting in
\begin{eqnarray}
    16t_{AA}+4(\Delta_{s}^2-v^2-\mu t_{AA}-4t_{AA}^2)\nonumber\\[3pt]
    -4\Delta_{s}\cos^2(\phi_{s})+(\mu+2t_{AA})^2=0\,,
\end{eqnarray}
which reduces to the minimal model solution (Eq.~(35) in the main text) for $t_{AA}=0$. The other transition points can be calculated in the same way as described for $\Delta_{p}$ with solutions $S_\pm$=0, where here
\begin{eqnarray}
    S_\pm\hspace{-0.3mm}\!&=&\!\frac{\!-2\Delta_s^2\!+\!8t_{AA}^2\!\!+\!4\mu t_{AA}\!+\!2v^2\!\pm\!2\sqrt{f}}{2\Delta_s^2\!+\!8t_{AA}^2\!-\!4\mu t_{AA}\!-\!2v^2\!\mp\!2\sqrt{f}}\nonumber\\[4pt]
    &&+\!\frac{\mu^2\!+\!4t_{AA}^2\!\!+\!4\mu t_{AA}}{\mu^{2}\!-\!4t_{AA}^2},
\end{eqnarray}
where
\begin{eqnarray}
    f &=& (v^2-\Delta_s^2)(-2\Delta_s^2+8t_{AA}^2+4\mu t_{AA}+2v^2)\nonumber\\
    &&+16\Delta_s^2t_{AA}^2\cos^2\phi_s\,.
\end{eqnarray}
Solutions to these equations that correspond to real $k$ values in the range $0\leq k<\pi$ correspond to points of topological phase transition between phases $N_\mathrm{D}^\mathrm{NS}=2$ and $N_\mathrm{D}^\mathrm{NS}=4$, and between phases $N_\mathrm{D}^\mathrm{NS}=1$ and $N_\mathrm{D}^\mathrm{NS}=3$ in the $\mathbb{Z}_4$ model. In the limit $t_{AA}\rightarrow0$, these solutions approach the minimal model transitions of $\mu=0$ and $\cos\phi_s=0$, depending on whether the Majorana number is 0 or 1. The determining factor on which of the two equations are used is identical to the case for the inclusion of $\Delta_p$ pairing, Eq.~(\ref{Sapproachcondition}). We note that an identical solution can be obtained through the integral calculation described in Ref.~\cite{tymczyszyn24}, by tracking the path of the integral and calculating at which point it first crosses the discontinuity at the origin.

\section{Topolectric circuit for the $\mathbb{Z}_4$ model with long-range parameters}

While we do not explicitly take into account the effects of long-range parameters on the impedance spectra, for completeness we consider here the possible construction of these components as an RLC circuit. A schematic of the circuit representing long-range parameters only, i.e., excluding the parameters shown in Fig.~\ref{topoz4schematic}, is shown in Fig.~\ref{z4longrangetopo}. The parameter $\Delta_p$ is represented by constant inter-channel inductors $L_P$ and capacitors $C_P$, while the parameter $t_{AA}$ is represented by the addition of capacitors $C_{AA}$ to the electron channel and inductors $L_{AA}$ to the hole channel. This results in additional constraints on the resonant frequency, Eq.~(\ref{z4frequency}), with
\begin{eqnarray}
    \omega_0&\equiv& 1/\sqrt{L_0C_0} = 1/\sqrt{L_1C_1}=1/\sqrt{L_SC_S}\nonumber\\
    &&=1/\sqrt{L_{AA}C_{AA}}=1/\sqrt{L_PC_P}\,.
    \label{hofreq}
\end{eqnarray} 
At the resonant frequency the parameterizations of Hamiltonian~(\ref{hoham}) in terms of circuit components are
\begin{eqnarray}
    \mu &=& -2C_1-2C_{AA}+C_0\,, \\
    v &=& -C_1 \,,\\
    t_{AA} &=& -C_{AA}\\
    \Delta_s &=& C_S^2+L_SC_S/R_S^2\,, \\[1.5pt]
    \phi_s &=& \arg\left[\eta_tC_S+i\sqrt{L_SC_S}/R_S\right]\,,\\
    \Delta_p &=& -C_P\,,
\end{eqnarray}
where we have preferentially represented the parameters in terms of capacitors, representations preferring inductors can be related by the resonant frequency, Eq.~(\ref{hofreq}).


\begin{thebibliography}{99}



\bibitem{kitaev01}
A. Y. Kitaev, Unpaired Majorana fermions in quantum wires, Phys.-Usp.\ {\bf 44}, 131 (2001).

\bibitem{kitaev03}
A. Y. Kitaev, Fault-tolerant quantum computation by anyons,
Ann.\ Phys.\ (N. Y), {\bf 303}, 2 (2003).

\bibitem{sarma06}
S. Das Sarma, M. Freedman, and C. Nayak, Topological quantum computation, 
Phys.\ Today {\bf 59} (7), 32 (2006).

\bibitem{nayak08}
C. Nayak, S. H. Simon, A. Stern, M. Freedman, and S. Das. Sarma, Non-Abelian anyonsand topological quantum computation,
Rev.\ Mod.\ Phys.\ {\bf 80}, 1083 (2008).

\bibitem{stern13}
A. Stern, and N. H. Lindner, Topological quantum computation—from basic concepts to first experiments, Sci.\ {\bf 339}, 1179 (2013).

\bibitem{lahtinen17}
V. Lahtinen, and J. K. Pachos, A short introduction to topological quantum computation,
SciPost Phys.\ {\bf 3}, (2017).

\bibitem{beenakker15}
C. W. J. Beenakker, 
Random-matrix theory of Majorana fermions and topological superconductors,
Rev. Mod. Phys. {\bf 87}, 1037 (2015).

\bibitem{guo16}
H.-M. Guo,
A brief review on one-dimensional topological insulators and superconductors,
Sci.\ China Phys.\ Mech.\ {\bf 59}, 637401 (2016).

\bibitem{sato17}
M. Sato and Y. Ando,
Topological superconductors: A review,
Rep.\ Prog.\ Phys.\ {\bf 80}, 076501 (2017).

\bibitem{sharma22}
M. M. Sharma, P. Sharma, N. K. Karn, and V. P. S. Awana, Comprehensive review on topological superconducting materials and interfaces,
Supercond.\ Sci.\ Technol.\ {\bf 35}, 083003 (2022)

\bibitem{dvir23}
T. Dvir, G. Wang, N. van Loo, C.-X. Liu, G. P. Mazur, A. Bordin, S. L. D. ten Haaf, J.-Y. Wang, D. van Driel, F. Zatelli, et al., Realization of a minimal Kitaev chain in coupled quantum dots, 
Nature {\bf 614}, 445 (2023).

%\bibitem{budich13}
%J. C. Budich and E. Ardonne,
%Equivalent topological invariants for one-dimensional Majorana wires in symmetry class D,
%Phys.\ Rev.\ B {\bf 88}, 075419 (2013).

\bibitem{laubscher24}
K. Laubscher, J. D. Sau, and S. D. Sarma, Majorana zero modes in gate-defined germanium hole nanowires,
Phys.\ Rev.\ B {\bf 109}, 035433 (2024)







\bibitem{mourik12}
V. Mourik, K. Zuo, S. M. Frolov, S. R. Plissard, E. P. A. M. Bakkers, and L. P. Kouwenhoven, Signatures of Majorana fermions in hybrid superconductor-semiconductor nanowire devices, 
Sci. {\bf 336}, 1003 (2012).

\bibitem{das12}
A. Das, Y. Ronen, Y. Most, Y. Oreg, M. Heiblum, and H. Shtrikman, Zero-bias peaks and splitting in an Al–InAs nanowire topological superconductor as a signature of Majorana fermions,
Nat.\ Phys.\ {\bf 8}, 887 (2012).

\bibitem{deng16}
M. T. Deng, S. Vaitiekėnas, E. B. Hansen, J. Danon, M. Leijnse, K. Flensberg, J. Nygård, P. Krogstrup, and C. M. Marcus, Majorana bound state in a coupled quantum-dot hybrid-nanowire system,
Sci.\ {\bf 354}, 1557 (2016).

\bibitem{nadj14}
S. Nadj-Perge, I. K. Drozdov, J. Li, H. Chen, S. Jeon, J. Seo, A. H. MacDonald, B. A. Bernevig, and A. Yazdani, Observation of Majorana fermions in ferromagnetic atomic chains on a superconductor,
Sci.\ {\bf 346}, 602 (2014).

\bibitem{ruby15}
M. Ruby, F. Pientka, Y. Peng, F. von Oppen, B. W. Heinrich, and K. J. Franke, End States and Subgap Structure in Proximity-Coupled Chains of Magnetic Adatoms,
Phys.\ Rev.\ Lett.\ {\bf 115}, 197204 (2015).

\bibitem{awlak16}
R. Awlak, M. Kisiel, J. Klinovaja, T. Meier, S. Kawai, T. Glatzel, D. Loss, and E. Meyer, Probing atomic structure and Majorana wavefunctions in mono-atomic Fe chains on superconducting Pb surface,
npj Quant.\ Inf.\ {\bf 2}, 16035 (2016). 

\bibitem{jack21}
B. Jäck, Y. Xie, and A. Yazdani, Detecting and distinguishing Majorana zero modes with the scanning tunnelling microscope. 
Nat.\ Rev.\ Phys.\ {\bf 3}, 541 (2021).

\bibitem{chen11}
X. Chen, Z.-C. Gu, and X.-G. Wen, Classification of gapped symmetric phases in one-dimensional spin systems, Phys.\ Rev.\ B {\bf 83}, 035107 (2011).

\bibitem{fidkowski11}
L. Fidkowski and A. Kitaev, Topological phases of fermions in one dimension, Phys.\ Rev.\ B {\bf 83}, 075103 (2011).

\bibitem{senthil15}
T. Senthil, Symmetry protected topological phases of quantum matter, Annu.\ Rev.\ Condens.\ Matter Phys. {\bf 6}, 299 (2015).

\bibitem{lutchyn18}
R. M. Lutchyn, E. P. A. M. Bakkers, L. P. Kouwenhoven, P. Krogstrup, C. M. Marcus, and Y. Oreg, Majorana zero modes in superconductor-semiconductor heterostructures, Nat.\ Rev.\ Mater. {\bf 3}, 52 (2018).

\bibitem{lee18}
C. H. Lee, S. Imhof, C. Berger, F. Bayer, J. Brehm, L. W. Molenkamp, T. Kiessling, and R. Thomale, Topolectrical circuits, Commun.\ Phys.\ {\bf 1}, 39 (2018).

\bibitem{imhof18}
S. Imhof, C. Berger, F. Bayer, J. Brehm, L. W. Molenkamp, T. Kiessling, F. Schindler, C. H. Lee, M. Greiter, T. Neupert, and R. Thomale, Topolectrical-circuit realization of topological corner modes, Nat.\ Phys.\ {\bf 14}, 925 (2018).

\bibitem{ezawa18}
M. Ezawa, Higher-order topological electric circuits and topological corner resonance on the breathing kagome and pyrochlore lattices, Phys.\ Rev.\ B {\bf 98}, 201402(R) (2018).

\bibitem{hofmann19}
T. Hofmann, T. Helbig, C. H. Lee, M. Greiter, and R. Thomale, Chiral Voltage Propagation and Calibration in a Topolectrical Chern Circuit, Phys.\ Rev.\ Lett.\ {\bf 122}, 247702 (2019).

\bibitem{ezawa19}
M. Ezawa, Braiding of Majorana-like corner states in electric circuits and its non-Hermitian generalization, Phys.\ Rev.\ B {\bf 100}, 045407 (2019).

\bibitem{liu20}
S. Liu, S. Ma, C. Yang, L. Zhang, W. Gao, Y. J. Xiang, T. J. Cui, and S. Zhang, Gain- and loss-induced topological insulating phase in a non-Hermitian electrical circuit, Phys.\ Rev.\ Appl.\ {\bf 13}, 014047 (2020).

\bibitem{yang20}
H. Yang, Z.-X. Li, Y. Liu, Y. Cao, and P. Yan, Observation of symmetry-protected zero modes in topolectrical circuits, Phys.\ Rev.\ Research {\bf 2}, 022028(R) (2020).

\bibitem{helbig20}
T. Helbig, T. Hofmann, S. Imhof, M. Abdelghany, T. Kiessling, L. W. Molenkamp, C. H. Lee, A. Szameit, M. Greiter and R. Thomale, Generalized bulk–boundary correspondence in non-Hermitian topolectrical circuits, Nat.\ Phys.\ {\bf 16}, 747 (2020).

\bibitem{ezawa20}
M. Ezawa, Non-Abelian braiding of Majorana-like edge states and topological quantum computations in electric circuits, Phys.\ Rev.\ B {\bf 102}, 075424 (2020).

\bibitem{wu20}
J. Wu, X. Huang, J. Lu, Y. Wu, W. Deng, F. Li, and Z. Liu, Observation of corner states in second-order topological electric circuits, Phys.\ Rev.\ B {\bf 102}, 104109 (2020).

\bibitem{dong21}
J. Dong, V. Juričić, and B. Roy, 
Topolectric circuits: Theory and construction,
Phys.\ Rev.\ Research {\bf 3}, 023056 (2021).

\bibitem{haydar23}
H. Sahin, Z. B. Siu, S. M. Rafi-Ul-Islam, J. F. Kong, M. B. A. Jalil, and C. H. Lee, Impedance responses and size-dependent resonances in topolectrical circuits via the method of images, 
Phys.\ Rev.\ B {\bf 107}, 245114 (2023).

\bibitem{haydar25}
H. Sahin, M. B. A. Jalil, and C. H. Lee, Topolectrical circuits—Recent experimental advances and developments, 
APL Electron.\ Devices {\bf 1}, 021503 (2025).

\bibitem{tang25}
Y. Tang, J. Wu, P. Lai, Y. Hu, H. Liu, W. Deng, H. Cheng, Z. Liu, and S. Chen, Loss induced delocalization of topological boundary modes. Nat.\ Commun.\ {\bf 16}, 10420 (2025).

\bibitem{chiu16}
C. Chiu, J. C. Y. Teo, A. P. Schnyder, and S. Ryu, Classification of topological quantum matter with symmetries
Rev.\ Mod.\ Phys.\ {\bf 88}, 035005 (2016).

\bibitem{altland97}
A. Altland and M. R. Zirnbauer, Nonstandard symmetry classes in mesoscopic normal-superconducting hybrid structures,
Phys.\ Rev.\ B {\bf 55}, 1142 (1997).

\bibitem{schnyder08}
A. P. Schnyder, S. Ryu, A. Furusaki, and A. W. W. Ludwig, Classification of topological insulators and superconductors in three spatial dimensions,
Phys.\ Rev.\ B {\bf 78}, 195125 (2008)

\bibitem{kitaev09}
A. Kitaev,
Periodic table for topological insulators and superconductors,
AIP Conf.\ Proc.\ {\bf 1134}, 22 (2009).

\bibitem{qi10}
X. L. Qi, Taylor, L. Hughes, and S. C. Zhang, Topological invariants for the Fermi surface of a time-reversal-invariant superconductor, Phys.\ Rev.\ B {\bf 81}, 134508 (2010).

\bibitem{ryu10}
S. Ryu, A. P. Schnyder, A. Furusaki, and A. W. W. Ludwig, Topological insulators and superconductors: tenfold way and dimensional hierarchy, New J.\ Phys.\ {\bf 12}, 065010 (2010).

\bibitem{kane10}
J. C. Y. Teo and C. L. Kane, Topological defects and gapless modes in insulators and superconductors, Phys.\ Rev.\ B {\bf 82}, 115120 (2010).

\bibitem{matveeva23}
P. Matveeva, T. Hewitt, D. Liu, K. Reddy, D. Gutman, and S. T. Carr, One-dimensional noninteracting topological insulators with chiral symmetry
Phys.\ Rev.\ B {\bf 107}, 075422 (2023).

\bibitem{shiozaki16}
K. Shiozaki, M. Sato, and K. Gomi,
Topology of nonsymmorphic crystalline insulators and superconductors,
Phys.\ Rev.\ B {\bf 93}, 195413 (2016).

\bibitem{teo08}
J. C. Y. Teo, L. Fu, and C. L. Kane, Surface states and topological invariants in three-dimensional topological insulators: Application to $\mathrm{Bi}_{1-x}\mathrm{Sb}_{x}$,
Phys.\ Rev.\ B {\bf 78}, 045426 (2008).

\bibitem{fu11}
L. Fu, Topological Crystalline Insulators
Phys.\ Rev.\ Lett.\ {\bf 106}, 106802 (2011).

\bibitem{hsieh12}
T. H. Hsieh, H. Lin, J. Liu, W. Duan, A. Bansil, and L. Fu, Topological crystalline insulators in the SnTe material class
Nat.\ Commun.\ {\bf 3}, 982 (2012).

\bibitem{liu14}
C.-X. Liu, R.-X. Zhang, and B. K. VanLeeuwen,
Topological nonsymmorphic crystalline insulators,
Phys.\ Rev.\ B {\bf 90}, 085304 (2014).

\bibitem{shiozaki14}
K. Shiozaki and M. Sato,
Topology of crystalline insulators and superconductors,
Phys.\ Rev.\ B {\bf 90}, 165114 (2014).

\bibitem{young15}
S. M. Young and C. L. Kane,
Dirac semimetals in two dimensions,
Phys.\ Rev.\ Lett.\ {\bf 115}, 126803 (2015).

\bibitem{wang16}
Z. Wang, A. Alexandradinata, R. J. Cava, and B. A. Bernevig,
Hourglass fermions,
Nature {\bf 532}, 189 (2016).


\bibitem{varjas17}
D. Varjas, F. de Juan, and Y.-M. Lu,
Space group constraints on weak indices in topological insulators,
Phys.\ Rev.\ B {\bf 96}, 035115 (2017).

\bibitem{kruthoff17}
J. Kruthoff, J. de Boer, J. van Wezel, C. L. Kane, and R.-J. Slager,
Topological classification of crystalline insulators through band structure combinatorics,
Phys.\ Rev.\ X {\bf 7}, 041069 (2017).

\bibitem{herrera22}
M. A. J. Herrera and D. Bercioux,
Tunable Dirac points in a two-dimensional nonsymmorphic wallpaper group lattice,
Commun.\ Phys.\ {\bf 6}, 42 (2023).

\bibitem{cayssol21}
J. Cayssol and J.-N. Fuchs, Topological and geometrical aspects of band theory, J.\ Phys.\ Mater. \ {\bf 4}, 034007 (2021).

\bibitem{su79}
W. P. Su, J. R. Schrieffer, and A. J. Heeger, Solitons in Polyacetylene, Phys.\ Rev.\ Lett.\ {\bf 42}, 1698 (1979).

\bibitem{su80}
W. P. Su, J. R. Schrieffer, and A. J. Heeger, Soliton excitations in polyacetylene, Phys.\ Rev.\ B 2{\bf 2}, 2099 (1980).

\bibitem{hasan10}
M. Z. Hasan and C. L. Kane, Colloquium: Topological insulators, 
Rev.\ Mod.\ Phys.\ {\bf 82}, 3045 (2010).

\bibitem{asboth16}
J. K. Asbóth, L. Oroszlány, and A. Pályi, A Short Course on Topological Insulators (Springer, Cham, 2016).

\bibitem{shiozaki15}
K. Shiozaki, M. Sato, and K. Gomi,
$\mathbb{Z}_2$ topology in nonsymmorphic crystalline insulators: M\"obius twist in surface states,
Phys.\ Rev.\ B {\bf 91}, 155120 (2015).

\bibitem{brzezicki20}
W. Brzezicki and T. Hyart,
Topological domain wall states in a nonsymmorphic chiral chain,
Phys.\ Rev.\ B {\bf 101}, 235113 (2020).

\bibitem{allen22}
R. E. J. Allen, H. V. Gibbons, A. M. Sherlock, H. R. M. Stanfield, and E. McCann,
Nonsymmorphic chiral symmetry and solitons in the Rice-Mele model,
Phys.\ Rev.\ B {\bf 106}, 165409 (2022).

\bibitem{kivelson83}
S. Kivelson, Solitons with adjustable charge in a commensurate Peierls insulator, 
Phys.\ Rev.\ B {\bf 28}, 2653 (1983).

\bibitem{fuchs21}
J.-N. Fuchs and F. Piéchon, Orbital embedding and topology of one-dimensional two-band insulators, 
Phys.\ Rev.\ B {\bf 104}, 235428 (2021).

%\bibitem{rice82}
%M. J. Rice and E. J. Mele, Elementary Excitations of a Linearly Conjugated Diatomic Polymer,
%Phys.\ Rev.\ Lett.\ {\bf 49}, 1455 (1982).

\bibitem{mong10}
R. S. K. Mong, A. M. Essin, and J. E. Moore,
Antiferromagnetic topological insulators,
Phys. Rev. B {\bf 81}, 245209 (2010).

\bibitem{fang15}
C. Fang and L. Fu,
New classes of three-dimensional topological crystalline insulators: Nonsymmorphic and magnetic,
Phys.\ Rev.\ B {\bf 91}, 161105(R) (2015).

\bibitem{zhao16}
Y. X. Zhao and A. P. Schnyder, Nonsymmorphic symmetry-required band crossings in topological semimetals, Phys.\ Rev.\ B {\bf 94}, 195109 (2016).

\bibitem{yanase17}
Y. Yanase and K. Shiozaki,
M\"obius topological superconductivity in UPt$_3$,
Phys. Rev. B {\bf 95}, 224514 (2017).

\bibitem{arkinstall17}
J. Arkinstall, M. H. Teimourpour, L. Feng, R. El-Ganainy, and H. Schomerus,
Topological tight-binding models from nontrivial square roots,
Phys.\ Rev.\ B {\bf 95}, 165109 (2017).

\bibitem{otrokov19}
M. M. Otrokov, I. I. Klimovskikh, H. Bentmann, D. Estyunin, A. Zeugner, Z. S. Aliev, S. Gaß, A. U. B. Wolter, A. V. Koroleva, A. M. Shikin, {\it et al.},
Prediction and observation of an antiferromagnetic topological insulator,
Nature {\bf 576}, 416 (2019).

\bibitem{gong19}
Y. Gong, J. Guo, J. Li, K. Zhu, M. Liao, X. Liu, Q. Zhang, L. Gu, L. Tang, X. Feng, {\it et al.},
Experimental realization of an intrinsic magnetic topological insulator,
Chin.\ Phys.\ Lett.\ {\bf 36}, 076801 (2019).

\bibitem{zhang19}
D. Zhang, M. Shi, T. Zhu, D. Xing, H. Zhang, and J. Wang,
Topological axion states in the magnetic insulator MnBi$_2$Te$_4$ with the quantized magnetoelectric effect,
Phys.\ Rev.\ Lett.\ {\bf 122}, 206401 (2019).

\bibitem{niu20}
C. Niu, H. Wang, N. Mao, B. Huang, Y. Mokrousov, and Y. Dai,
Antiferromagnetic topological insulator with nonsymmorphic protection in two dimensions,
Phys.\ Rev.\ Lett.\ {\bf 124}, 066401 (2020).

\bibitem{marques19}
A. M. Marques and R. G. Dias,
One-dimensional topological insulators with noncentered inversion symmetry axis,
Phys.\ Rev.\ B {\bf 100}, 041104(R) (2019).



\bibitem{yang22}
Y. Yang, H. C. Po, V. Liu, J. D. Joannopoulos, L. Fu, and M. Solja\v{c}i\'{c},
Non-Abelian nonsymmorphic chiral symmetries,
Phys.\ Rev.\ B {\bf 106}, L161108 (2022).

\bibitem{mccann23}
E. McCann, Catalog of noninteracting tight-binding models with two energy bands in one dimension, Phys.\ Rev.\ B {\bf 107}, 245401 (2023).


\bibitem{tymczyszyn24}
M. Tymczyszyn and E. McCann, One-dimensional $\mathbb{Z}_{4}$ topological superconductor, Phys.\ Rev.\ B {\bf 110}, 085416 (2024)


\bibitem{tanaka12}
Y. Tanaka, M. Sato, and N. Nagaosa,
Symmetry and Topology in Superconductors - Odd-Frequency Pairing and Edge States -,
J.\ Phys.\ Soc.\ Jpn.\ {\bf 81}, 011013 (2012).

\bibitem{alicea12}
J. Alicea,
New directions in the pursuit of Majorana fermions in solid state systems,
Rep.\ Prog.\ Phys.\ {\bf 75}, 076501 (2012).

\bibitem{leijnse12}
M. Leijnse and K. Flensberg,
Introduction to topological superconductivity and Majorana fermions,
Semicond.\ Sci.\ Technol.\ {\bf 27}, 124003 (2012).

%\bibitem{dyson62}
%F. J. Dyson, Statistical Theory of the Energy Levels of Complex Systems. I, J.\ Math.\ Phys.\ {\bf 3}, 140 (1962). 

%\bibitem{wigner67}
%E. P. Wigner, Random matrices in physics, SIAM Rev.\ {\bf 9}, 1 (1967).

%\bibitem{mehta90}
%M. L. Mehta, Random Matrix Theory (Springer, New York, (1990).

%\bibitem{guhr98}
%T. Guhr, A. Müller-Groeling, and H. A. Weidenmüller,
%Random-matrix theories in quantum physics: Common concepts, Phys.\ Rep.\ {\bf 299}, 189 (1998).

%\bibitem{oganesyan07}
%V. Oganesyan and D. A. Huse, Localization of interacting fermions at high temperature, Phys.\ Rev.\ B {\bf 75}, 155111 (2007).

%\bibitem{atas13}
%Y. Y. Atas, E. Bogomolny, O. Giraud, and G. Roux, Distribution of the ratio of consecutive level spacings in random matrix ensembles, Phys.\ Rev.\ Lett.\ {\bf 110}, 084101 (2013).



%\bibitem{wong12}
%C. L. M. Wong and K. T. Law, Majorana Kramers doublets in $d_{x^{2}}-d_{y^{2}}$-wave superconductors with Rashba spin-orbit coupling, Phys.\ Rev.\ B {\bf 86}, 184516 (2012).

%\bibitem{schnyder11}
%A. P. Schnyder and S. Ryu, Topological phases and surface flat bands in superconductors without inversion symmetry, Phys.\ Rev.\ B {\bf 84}, 060504(R) (2011).

%\bibitem{ghol18}
%S. Gholizadeh, M. Yahyavi and B. Hetényi, Extended Creutz ladder with spin-orbit coupling: A one-dimensional analog of the Kane-Mele model, Europhys.\ Lett.\ {\bf 122}, 27001 (2018).

%\bibitem{supplementarymaterial}
%See Supplemental Material at [URL will be inserted by publisher] for long-range parameters of the generalized superconducting Rice-Mele model, example models of the symmorphic D, DIII, BDI, CII classes and nonsymmorphic CII class, and for the topological transition points of the $\mathbb{Z}_4$ model in the presence of A-A and B-B hopping and superconducting pairing.

%\bibitem{bena09}
%C. Bena and G. Montambaux, Remarks on the tight-binding model of graphene, New J.\ Phys.\ {\bf 11}, 095003 (2009).

\bibitem{cheng12}
M. Cheng, Superconducting proximity effect on the edge of fractional topological insulators, Phys.\ Rev.\ B {\bf 86}, 195126 (2012).

\bibitem{zhang14}
F. Zhang and C.L. Kane, Time-Reversal-Invariant ${Z}_4$ Fractional Josephson Effect, Phys.\ Rev.\ Lett.\ {\bf 113}, 036401 (2014).

\bibitem{cheong15}
S. Cheon, T. Kim, S. Lee, H. W. Yeom, Chiral solitons in a coupled double Peierls chain, Sci.\ {\bf 350}, 182-185 (2015).


\bibitem{jackiw76}
R. Jackiw and C. Rebbi, Solitons with fermion number ½, Phys.\ Rev.\ D {\bf 13}, 3398 (1976).

\bibitem{muten24}
J. H. Muten, L. H. Frankland, and E. McCann, Solitons in binary compounds with stacked two-dimensional honeycomb lattices, Phys.\ Rev.\ B {\bf 109}, 165416 (2024).

\bibitem{han20}
S.-H. Han, S.-G. Jeong, S.-W. Kim, T.-H. Kim, and S. Cheon,
Topological features of ground states and topological solitons
in generalized Su-Schrieffer-Heeger models using generalized
time-reversal, particle-hole, and chiral symmetries, Phys.\ Rev.\
B {\bf 102}, 235411 (2020).

\bibitem{koschny02}
T. Koschny and L. Schweitzer, Influence of correlated disorder potentials on the levitation of current carrying states in the quantum Hall effect, Physica\ E\ {\bf 12}, 654 (2002).

\bibitem{guo08}
Z.-Z. Guo, Entanglement in one-dimensional Anderson model with long-range correlated disorder, Chin.\ Phys.\ Lett.\ {\bf 25}, 1079
(2008).


\bibitem{franceschi10}
S. De Franceschi, L. Kouwenhoven, C. Schönenberger and W. Wernsdorfer, Hybrid superconductor–quantum dot devices,
Nat.\ Nanotechnol.\ {\bf 5}, 703 (2010).

\bibitem{albrecht16}
S. M. Albrecht, A. P. Higginbotham, M. Madsen, F. Kuemmeth, T. S. Jespersen, J. Nygård, P. Krogstrup and C. M. Marcus, Exponential protection of zero modes in Majorana islands,
Nat.\ {\bf 531}, 206 (2016).

\bibitem{nikolaenko21}
A. Nikolaenko and F. Pientka, Topological superconductivity in proximity to type-II superconductors,
Phys.\ Rev.\ B {\bf 103}, 134503 (2021).

\bibitem{data}
E. McCann and M. Tymczyszyn, Research data for published version of ``One-dimensional topology and topolectrics of nonsymmorphic Kramers degenerate systems". 
\url{https://doi.org/10.17635/lancaster/researchdata/741}.


\end{thebibliography}
\end{document}